\documentclass[10pt,journal,compsoc]{IEEEtran}
\ifCLASSOPTIONcompsoc
  \usepackage[nocompress]{cite}
\else
  \usepackage{cite}
\fi
\ifCLASSINFOpdf
\else
\fi
\usepackage{amsfonts,amsmath}
\usepackage{algpseudocode}
\usepackage{algorithmicx}
\usepackage{graphicx,epsfig}
\usepackage{color}

\usepackage{cite}

\usepackage[ruled]{algorithm2e}

\usepackage{subcaption}
\usepackage{tabularx} 

\def \_ {\rule{.3em}{.15ex}} 

 \normalmarginpar
\newcommand{\mymarginpar}[1]{\marginpar{#1}}
\renewcommand{\marginpar}[1]{}

\newcommand{\ls}[1]{\dimen 0= \fontdimen 6 \the \font
    \lineskip =#1 \dimen0
    \advance \lineskip .5 \fontdimen 5 \the \font
    \advance \lineskip - \dimen 0
    \lineskiplimit =.9 \lineskip
    \baselineskip = \lineskip
    \advance \baselineskip \dimen 0
    \normallineskip \lineskip
    \normallineskiplimit \lineskiplimit
    \normalbaselineskip \baselineskip
	\ignorespaces
}

\newcommand{\bearn}{\begin{eqnarray*}}
\newcommand{\eearn}{\end{eqnarray*}}
\newcommand{\barr}{\begin{array}}
\newcommand{\earr}{\end{array}}

\newcommand{\N}{{\cal N}}

\newtheorem{definition}{Definition}
\newtheorem{property}[definition]{Property}
\newtheorem{proposition}[definition]{Proposition}
\newtheorem{lemma}[definition]{Lemma}
\newtheorem{theorem}[definition]{Theorem}
\newtheorem{corollary}[definition]{Corollary}
\newtheorem{example}[definition]{Example}
\newtheorem{remark}[definition]{Remark}

\newcommand{\benum}{\begin{enumerate}}
\newcommand{\eenum}{\end{enumerate}}

\newcommand{\bdesc}{\begin{description}}
\newcommand{\edesc}{\end{description}}

\newcommand{\bfig}[2]{
	\begin{figure}\centering\includegraphics[width=#2]{#1}
	}
\newcommand{\brotatefig}[2]{
	\begin{figure}[htbp]\centerline{
			\epsfig{figure={#1},clip=,angle=-90,width={#2}}
		}
	}

\newcommand{\bfigfirst}[2]{\begin{figure}[h]\centerline{\setlength{\epsfxsize}{#2}\epsffile{#1}}}
\newcommand{\efig}[2]{\caption{#2}\label{fig:#1}\end{figure}\mymarginpar{fig:#1}}
\newcommand{\erotatefig}[2]{\caption{#2}\label{fig:#1}\end{figure}\mymarginpar{fig:#1}}
\newcommand{\rfig}[1]{Figure \ref{fig:#1}}
\newcommand{\btab}[1]{\begin{table}\centering\begin{tabular}{#1}}
\newcommand{\etab}[3]{\end{tabular}\caption[#3]{#2}\label{tab:#1}\end{table}\mymarginpar{tab:#1}\vspace{.1in}}

\newcommand{\btabular}[1]{\begin{center}\begin{tabular}{#1}}
\newcommand{\etabular}{\end{tabular}\end{center}}
\newcommand {\bdefin}[1]{\begin{definition}\mymarginpar{def:#1}\label{def:#1}}
\newcommand {\edefin}{\end{definition}}
\newcommand {\rdef}[1]{Definition \ref{def:#1}}
\newcommand {\bpro}[1]{\begin{property}\mymarginpar{pro:#1}\label{pro:#1}}
\newcommand {\epro}{\end{property}}

\newcommand {\bprop}[1]{\begin{proposition}\mymarginpar{prop:#1}\label{prop:#1}}
\newcommand {\eprop}{\end{proposition}}
\newcommand {\rprop}[1]{Proposition \ref{prop:#1}}
\newcommand {\blem}[1]{\begin{lemma}\mymarginpar{lem:#1}\label{lem:#1}}
\newcommand {\elem}{\end{lemma}}
\newcommand {\rlem}[1]{Lemma \ref{lem:#1}}
\newcommand {\bthe}[1]{\begin{theorem}\mymarginpar{the:#1}\label{the:#1}}
\newcommand {\ethe}{\end{theorem}}
\newcommand {\rthe}[1]{Theorem \ref{the:#1}}
\newcommand {\bproof}{\noindent {\bf Proof.} \ }
\newcommand {\eproof}{\hfill \squares \\ \vspace{.3cm}}
\newcommand {\bcor}[1]{\begin{corollary}\mymarginpar{cor:#1}\label{cor:#1}}
\newcommand {\ecor}{\end{corollary}}

\newcommand{\bax}[1]{\begin{axiom}\mymarginpar{ax:#1}\label{ax:#1}}
\newcommand{\eax}{\vspace{-.1in} \end{axiom}}

\newcommand{\bex}[2]{\vspace{.1in}\begin{example}\mymarginpar{ex:#1}{\bf #2}\label{ex:#1} }
\newcommand{\eex}{\end{example}\vspace{.3cm}}

\newcommand{\brem}[1]{\begin{remark}\mymarginpar{rem:#1}\label{rem:#1}\em}
\newcommand{\erem}{\end{remark}}

\newcommand{\beq}[1]{\mymarginpar{eq:#1}\begin{equation}\label{eq:#1}}
\newcommand{\beqno}[1]{\mymarginpar{eq:#1}\begin{eqnarray}\nonumber}
\newcommand{\eeq}{\end{equation}}
\newcommand{\eeqno}{&&\end{eqnarray}}
\newcommand{\req}[1]{(\ref{eq:#1})}

\newcommand{\bear}[1]{\mymarginpar{eq:#1}\begin{eqnarray}\label{eq:#1}}
\newcommand{\bearno}[1]{\mymarginpar{eq:#1}\begin{eqnarray}\nonumber}
\newcommand{\eear}{\end{eqnarray}}
\newcommand{\eearno}{\end{eqnarray}}
\usepackage{environ}
\NewEnviron{es}{%
	\begin{equation}\begin{split}
	\BODY
	\end{split}\end{equation}
}
\newcommand{\bsel}{\left\{\begin{array}{cl}}
\newcommand{\esel}{\end{array}\right.}
\newcommand{\bmat}[1]{\left[\begin{array}{#1}}
\newcommand{\emat}{\end{array}\right]}
\newcommand{\bsec}[2]{\mymarginpar{sec:#2}\section{#1}\label{sec:#2}}
\newcommand{\rsec}[1]{Section \ref{sec:#1}}

\newcommand{\bsubsec}[2]{\mymarginpar{sec:#2}\subsection{#1}\label{sec:#2}}

\def\R{I\kern-0.30em R}
\def\N{I\kern-0.30em N}
\def\P{I\kern-0.30em P}

\newcommand \squares{\vrule height6pt width7pt depth1pt}

\def \ex{{\bf\sf E}}
\def \pr{{\bf\sf P}}

\newcommand \rezprob{r}

\newcommand{\pon}{p_{1,1}^{(i)}}
\newcommand{\poff}{p_{0,0}^{(i)}}
\begin{document}

\title{ETTR Bounds and Approximation Solutions of Blind Rendezvous Policies in Cognitive Radio Networks with Random Channel States}
	
	

\author{Cheng-Shang~Chang,~\IEEEmembership{Fellow,~IEEE,}
                Duan-Shin Lee,~\IEEEmembership{Senior Member,~IEEE},\\
         Yu-Lun Lin, and Jen-Hung Wang
\IEEEcompsocitemizethanks{\IEEEcompsocthanksitem  C.-S. Chang,  D.-S. Lee, Y.-L. Lin and J.-H. Wang are with the Institute of Communications Engineering,
National Tsing Hua University,
Hsinchu 300, Taiwan, R.O.C.\protect\\
E-mail:   cschang@ee.nthu.edu.tw,  lds@cs.nthu.edu.tw, beck2245@gmail.com, f0960778676@gmail.com.
}
}

\markboth{IEEE TRANSACTIONS ON MOBILE COMPUTING,~Vol.~xx, No.~x, August~20xx}%
{Chang \MakeLowercase{\textit{et al.}}: Blind Rendezvous in Cognitive Radio Networks with Random Channel States}
%

\IEEEtitleabstractindextext{%
\begin{abstract}
In this paper, we consider the multichannel rendezvous problem in cognitive radio networks (CRNs) where the probability that
two users hopping on the same channel have a successful rendezvous  is a function of channel states.
The channel states are modelled by stochastic processes with  joint distributions known to users. However, the exact state of a channel at any time is not observable. We first consider two channel models: (i) the fast time-varying channel model (where the channel states are assumed to be independent and identically distributed in each time slot), and (ii) the slow time-varying channel model (where the channel states remain unchanged over time).
Among the classes of the  blind rendezvous policies that randomly hop on channels according to certain channel selection probabilities, we show
the optimal channel selection policy that minimizes the expected time-to-rendezvous (ETTR) is the single selection policy that  hops on the ``best'' channel all the time in the fast time-varying channel model. However, for the slow time-varying channel model, it is much more difficult to find the optimal channel selection policy. By using the majorization ordering, we derive  a lower bound and an upper bound for the ETTR under the assumption that the channel states are exchangeable random variables.
Bases on these bounds, we then prove various approximation solutions. We then extend our results to general channel models where
the joint distribution of the channel states is only assumed to be stationary in time.
\end{abstract}

}

\maketitle

\IEEEdisplaynontitleabstractindextext

%
\IEEEpeerreviewmaketitle

\IEEEraisesectionheading{\section{Introduction}\label{sec:introduction}}

%
%
%
%
\IEEEPARstart{T}{he}
multichannel rendezvous problem that asks two secondary users (SU) to find a common available channel (not used by primary users (PU)) has received a tremendous amount of attention in the research community of cognitive radio networks (CRNs) (see e.g., the tutorial and the book \cite{Tutorial,Book} and references therein). One simple solution  for the rendezvous problem, known as the {\em focal strategy} in \cite{AG03}, is for both users to select a
designated channel to meet. However, there are three well-known risks for doing that: (i) the designated channel may not be available to one of the two users, (ii) the designed channel might be congested as all the other users in the network come to this channel to meet \cite{wjliao,ToN2015}, and (iii) the designated channel is vulnerable to jamming attack by an adversary \cite{Krunz2015}.
As such, in the literature the multichannel rendezvous problem is generally solved by having each secondary user hopping on its available channels over time and both users are assumed to have a successful rendezvous when they both hop on a common available channel at the same time.
For such a   rendezvous problem, the objective is to minimize
the time-to-rendezvous (TTR), i.e., the first time that the two users have a successful rendezvous.

There are various {\em deterministic} channel hopping (CH) sequences proposed in the literature  that can guarantee
finite maximum time-to-rendezvous (MTTR) under various assumptions for CRNs, e.g., QCH \cite{Quorum}, DRSEQ \cite{DRSEQ}, Modular Clock \cite{Theis2011}, JS \cite{JS2011}, DRDS \cite{DRDS13}, FRCH \cite{ChangGY13}, ARCH \cite{ARCH}, CBH \cite{CBH2014}, and Two-prime Modular Clock \cite{ToN2017}. These {\em deterministic} CH algorithms, in general,  can be categorized by the following assumptions:
(i) the {\em symmetric} (resp. {\em asymmetric}) assumption in which users follow the same (resp. different) algorithm to generate their CH sequences, (ii) the {\em anonymous}  assumption in which users do not use their identifiers (ID), (iii) the {\em asynchronous} (resp. {\em synchronous}) assumption in which the clocks of users are not synchronized (resp. synchronized), (iv) the {\em heterogeneous} (resp. {\em homogeneous}) assumption in which users may perceive different (resp. the same) sets of available channels,  (v) the {\em multiple radio} assumption in which users may be equipped with multiple radios, and (vi) the {\em oblivious} (resp. {\em non-oblivious}) assumption in which the channel labels of users may be different (resp. are same). Two main mathematical theories behind these CH algorithms are the relative difference sets \cite{Quorum,DRDS13} (mostly for the homogeneous setting with a large number of common available channels) and the Chinese remainder theorem \cite{Theis2011,ToN2017} (mostly for the heterogeneous setting with a small number of common available channels).
There are some recent results (see, e.g., \cite{Multiradio14,RPS,AMRR,TCCN2018}) that considered the most challenging setting under the assumptions in (i)-(v). In particular, it was shown in
\cite{TCCN2018} that the MTTR in a CRN with $N$ commonly labelled channels can be bounded above
by $9 M \lceil n_1/m_1 \rceil \cdot \lceil n_2/m_2 \rceil$ time slots, where $n_1$ (resp. $n_2$) is the number of available channels to user $1$ (resp. 2), $m_1$ (resp. $m_2$) is the number of radios for user $1$ (resp. 2), and $M=2\lceil \log_2 (\lceil \log_2 N \rceil)\rceil+7$.
For more detailed descriptions of these {\em deterministic} CH algorithms and their assumptions, we refer to the book \cite{Book} and
the tutorial \cite{Tutorial}.

Though these {\em deterministic} CH algorithms with MTTR bounds are mostly elegant and beautiful in theory, they may not be practical for industrial use due to the following reasons:

\noindent (ii) As pointed out in \cite{ToN2017}, the simple blind rendezvous (random) algorithm is nearly optimal in terms of the expected time-to-rendezvous (ETTR)  and most of the rendezvous algorithms in the literature perform rather poorly in ETTR when compared to the simple blind rendezvous (random) algorithm.
The rationale behind that is because there is usually a “stay” mode in these CH algorithms and a user in its “stay” mode stays on the same channel for a rather long period of time.  When two users in the “stay” mode  staying on two different channels for a long period of time, a lot of time is wasted and that results in poor performance of ETTR.

\noindent (ii)  In practice, two secondary users might not have a successful rendezvous even when they both hop on a common available channel at the same time. This might due to several reasons, e.g., interferences from other secondary users in a heavily loaded channel (congestion of a channel) or degrading of signals due to channel fading.


In view of these, it is thus of importance to investigate the effects of random channel states on the ETTRs of rendezvous algorithms.
Though there are a lot of prior works on the multichannel rendezvous problem in CRNs,
it seems that there are only a very small number of papers that addressed the effect of random channel states in the literature.
In particular, Pu \emph{et al.} \cite{Pu2016} considered the channel state model in which there are only two channel states: available or unavailable (used by a PU).
 The probability that a channel is available to a user in a  time slot is chosen from a uniform distribution. Under such a channel state model with $N$ channels,
they proposed efficient algorithms that guarantee rendezvous for both synchronous and asynchronous users in $O(\log^2 N)$ and $O(\log^3 N)$ time slots with high probability respectively.
Al-Mqdashi \emph{et. al} \cite{Abdu2017} considered a more sophisticated channel state model that is characterized by a
three-state continuous-time Markov chain.
The three channel states are idle, PU occupied, or SU occupied.
For such a channel model, they proposed nested cyclic quorum channel hopping (NCQ-CH) and minimal nested cyclic quorum channel hopping (MNCQ-CH) to cope with the fast PU dynamics.

To take the channel state into account, we consider a more general
model than the two-state model in \cite{Pu2016} and the three-state model in \cite{Abdu2017}. In our model, each channel has several random states
and the probability that two secondary users hopping on a common channel have a successful rendezvous is a function of the channel state.
For a CRN with $N$ channels, the states of the $N$ channels are characterized by
the stochastic process $\{\boldsymbol{X}(t)=(X_1(t),X_2(t),\ldots,X_N(t)), t \ge 0\}$, where $X_i(t)$, $i=1,2, \ldots, N$, is the  random variable that represents the state of channel $i$ at time $t$.
When two secondary users hopping on a channel in state $x$, they will rendezvous with probability $\rezprob(x)$.
Since the event that two users hopping on a channel do not rendezvous has a nonzero probability, the MTTR cannot be bounded by a finite constant in our model. As such, MTTR
is not suitable for measuring the performance of rendezvous algorithms in our model.
Instead, we will use  ETTR as the performance metric.
Also, we assume that the exact state of a channel at any time is not observable by a user. The reason behind that  is because it is in general difficult for a user to know the congestion level of a channel (the number of users in a channel).
Since we are interested in the performance of ETTR and the channel states are not observable, we  limit ourselves to  the class of {\em blind rendezvous policies} in which
 each user selects channel $i$ {\em independently} with probability $p_i$, $i=1,2, \ldots, N$, in every time slot.
 Our objective is then to find the channel selection probabilities $p_i$'s so as to minimize the ETTR under our stochastic models of channel states. Note that such a  class of blind rendezvous policies can be easily implemented in the {\em symmetric, anonymous, asynchronous, homogeneous} and {\em non-oblivious} setting.

One natural question is whether there exists a universally optimal channel selection policy.
To address such a question, we consider two extreme models for the channel states: (i) the fast time-varying channel model and (ii) the slow time-varying channel model. In the fast time-varying channel model, the channel states change very fast and are assumed to be {\em independent} and {\em identically distributed} in each time slot (as in \cite{Pu2016}).
On the other hand, in the slow time-varying channel model, the channel states change very slowly and are assumed to {\em remain unchanged} during the rendezvous process. Since the joint distribution of the random channel states is known to each user, intuitively each user can compute the expected rendezvous probabilities of the $N$ channels and
chooses the ``best'' channel all the time to speed up the rendezvous process.
Such a single selection policy is indeed  the optimal policy
 for the fast time-varying channel model (see \rsec{fast_chmod} for the detailed proof).
 However, such an intuitive argument is no longer valid for the slow time-varying channel model.
This is because the ``best'' channel that has the largest expected rendezvous probability might have a nonzero probability to be in a very bad state with an extremely small rendezvous probability. As the states remain unchanged during the rendezvous process in the slow time-varying channel model, the single selection policy could lead to a very large ETTR if each user selects the ``best'' channel all the time.

Finding the optimal blind rendezvous policy for the slow time-varying channel model
 is in general very difficult, even when the $N$ channel state random variables,
$X_1(0),X_2(0),\ldots,X_N(0)$, are independent and identically distributed.
As such, we look for approximation solutions  for the slow time-varying channel model.
Our main results to the slow time-varying channel model are as follows:


\noindent (i)
Under the assumption that the states of the $N$ channels are {\em exchangeable} random variables,  we show by using the majorization ordering \cite{MarOl} that the ETTR of a blind rendezvous policy can be written as a product of two functions: one is a Schur concave function (that can be minimized by using the {\em single selection policy}) and the other is a Schur convex function (that can be minimized by using the {\em uniform selection policy}). This leads to a lower bound and an upper bound for the ETTR. The uniform selection policy (that selects each channel with an equal probability) is an $N$-approximation policy, i.e.,  the ETTR of the approximation policy is not greater than the $N$ times of the optimal policy.
On the other hand, the single selection policy is an $M$-approximation policy, where $M={\ex[\frac{1}{\rezprob({X}_1(0))} ]}/{\frac{1}{\ex[{\rezprob({X}_1(0))} ]}}$.

\noindent (ii) Under the assumption that the states of the $N$ channels are {\em independent and identically distributed} (i.i.d.) random variables with
only two states (state 0 as the bad state and state 1 as the good state), we prove two asymptotic lower bounds for the ETTR  in the asymptotic regime when $r(0) \to r(1)$ and $r(0) \to 0$. Based on these two asymptotic lower bounds, we
derive an asymptotic $(1+\epsilon)$-approximation solution for such a two-state channel model for $0 < \epsilon \le 3$.
The asymptotic $(1+\epsilon)$-approximation solution leads to a local search algorithm, called the
{\em improved uniform selection policy} in Algorithm \ref{alg:multipleb}.
 Various numerical experiments are conducted to show the effectiveness of Algorithm \ref{alg:multipleb}.

Finding the optimal blind rendezvous policy for  the general time-varying channel model is even much more difficult.
When the sequence of random vectors
$\{\boldsymbol{X}(t)=(X_1(t),X_2(t),\ldots, X_N(t)), t\ge 0\}$ are only assumed to be {\em stationary}, i.e.,
its joint distribution is invariant with respect to any time shift,
we show that its ETTR is upper bounded by the ETTR of the slow time-varying channel model.
In other words, the ETTR of the slow time-varying channel model serves as the worst case when we do not know the complete statistics of
the channel states.
On the other hand, we also generalize the slow time-varying two-state model to a two-state Markov chain model.
 We show that if the two-state Markov chain is positively correlated, then
its ETTR  is lower bounded by the ETTR of the  fast time-varying channel model.
As such, the ETTR of the  fast time-varying channel model is the best case when
we do not know the transition probabilities of
the positively correlated two-state Markov chain.
Based on both the lower bound and the upper bound, we further show that
the uniform selection policy is an asymptotic $N$-approximation solution and the
single selection policy is an $M$-approximation solution for any positively correlated two-state Markov chains.


The rest of the paper is organized as follows. We first describe the system model in \rsec{system}.
In \rsec{fast_chmod}, we introduce the fast  time-varying channel model and show that the optimal policy is the single selection policy.
Then we introduce the slow time-varying channel model  in \rsec{slow_chmod}, where we
first  use a change of probability vectors in \rsec{change} and the majorization ordering \rsec{lower} to derive bounds for the ETTR.
These bounds are then used for proving approximation algorithms in \rsec{solutions}.
We then consider the slow time-varying channel model with two states in \rsec{two}. There we show
two asymptotic lower bounds for the ETTR and the asymptotic $(1+\epsilon)$-approximation solution.
In \rsec{general}, we introduce a general time-varying channel model and derive its ETTR upper bound.
We then consider a Markov channel model with two states in \rsec{gtwo}, where we derive its ETTR lower bound.
The paper is then concluded in \rsec{conclusion}.

\bsec{System model}{system}

In this paper, we consider  a cognitive radio network (CRN) with $N$ channels (with $N \geq 2$), indexed from $1$ to $N$, in the
discrete-time setting where time is slotted  and indexed from $t=0,1,2,\ldots$.
We assume that there are $L$-states for each channel, indexed from $0,1,2,\ldots, L-1$.
Denote by $\rezprob(x)$ the rendezvous probability when  a channel in state $x$.  Then when two users hop on a channel in state $x$ at the same time, these two users will rendezvous  with probability $\rezprob(x)$, and this is independent of everything else.
Without loss of generality, we may order the $L$ channel states so that
 $\rezprob(x)$ is an increasing function of $x$, i.e.,
$$
		\rezprob(0) \le \rezprob(1) \le \rezprob(2) \le \ldots \le \rezprob(L-1).
$$

The states of the $N$ channels are characterized by
the stochastic process $\{\boldsymbol{X}(t)=(X_1(t),X_2(t),\ldots,X_N(t)), t \ge 0\}$, where $X_i(t)$, $i=1,2, \ldots, N$, is the  random variable that represents the state of channel $i$ at time $t$. We assume that the joint distribution of the stochastic process $\{\boldsymbol{X}(t), t \ge 0\}$ is known to each user. However, the exact state of a channel at any time is not observable by a user. The reason that we assume the exact state of a channel is not observable is because it is in general difficult for a user to know the congestion level of a channel (the number of users in a channel).


We consider  the class of {\em blind rendezvous policies}, i.e., at the $t^{th}$ time slot each user selects channel $i$ with probability $p_i$, $i=1,2, \ldots, N$. Such a channel selection is independent of everything else.
Suppose that the channel state of the $i^{th}$ channel at time $t$ is $x_i$, $i=1,,2 \ldots, N$.
Then under the blind rendezvous policy, the probability that these two users will have a successful rendezvous at time $t$ on channel $i$ is simply $p_i^2 \cdot \rezprob(x_i)$. This is because the two users have to hop on channel $i$ at time $t$ and the rendezvous is successful on channel $i$ with probability $\rezprob(x_i)$. As such, the two users will have a successful rendezvous at time $t$ is $\sum_{i=1}^N p_i^2  \cdot \rezprob(x_i)$.

In this paper, we will address the problem of finding a blind rendezvous policy (and the corresponding channel selection probabilities) that minimizes the expected time-to-rendezvous (ETTR). Specifically, we consider the following optimization problem that minimizes the ETTR  among the class of blind rendezvous policies:
\bear{opti0011}
&\min_{\boldsymbol{p}} \quad &\ex[T(\boldsymbol{p})]\nonumber \\
&s.t. \quad &p_i \ge 0,\; i=1,2, \ldots, N, \nonumber \\
&&\sum_{i=1}^N p_i=1,
\eear
where $T(\boldsymbol{p})$ is the time-to-rendezvous for the blind rendezvous policy with the channel selection probabilities
$\boldsymbol{p}=(p_1,p_2, \ldots, p_N)$. In this paper, we are particularly interested in two policies: (i) the {\em single selection policy} with
$p_1=1$ and $p_i=0$, $i=2, \ldots, N$, and (ii) the {\em uniform selection policy} with $p_i=1/N$, $i=1,2, \ldots, N$.

\bsec{The Fast Time-Varying Channel Model}{fast_chmod}

In the literature,  a fast time-varying channel is commonly referred to a channel whose channel state changes fast with respect to time.
In this regard, we define the fast time-varying channel model if the channel states are independent and identically distributed (i.i.d.) with respect to time. This is formally stated as follows:

\bdefin{fast}{\bf (Fast Time-Varying Channel Model)}
For the fast time-varying channel model, the sequence of random vectors
$\{\boldsymbol{X}(t)=(X_1(t),X_2(t),\ldots, X_N(t)), t\ge 0\}$ are assumed to be i.i.d. with the joint probability mass function
$$\pr(\boldsymbol{X}(0)=\boldsymbol{x})=q(\boldsymbol{x}),$$
where $\boldsymbol{x}=(x_1, x_2, \ldots, x_N)$.
\edefin

In the following theorem, we show that the single selection policy is the optimal policy in the fast time-varying channel model.

\bthe{fast}
Consider the fast time-varying channel model in \rdef{fast}. Suppose that
\beq{fast1111}
\ex[\rezprob(X_1(0))] \ge \ex[\rezprob(X_2(0))] \ge \ldots \ge \ex[\rezprob(X_N(0))].
\eeq
Then the single selection policy minimizes the ETTR in \req{opti0011} among the class of blind rendezvous policies.
\ethe

\bproof
From the i.i.d. assumption in the fast time-varying channel model, we know that the random variable $T(\boldsymbol{p})$ is geometrically distributed with parameter $\ex[\sum_{i=1}^{N}p_i^2\rezprob(X_i(0))]$. Thus, the ETTR is simply
$$\ex[{T(\boldsymbol{p})}]= 1 / \ex[\sum_{i=1}^{N}p_i^2\rezprob(X_i(0))].$$
Since the expectation operator is linear, we have
\beq{fastETTR}
\ex[{T(\boldsymbol{p})}]= 1 / \sum_{i=1}^{N}p_i^2 \ex[\rezprob(X_i(0))].
\eeq
Clearly,  minimizing the ETTR is equivalent to maximizing $\sum_{i=1}^{N}p_i^2 \ex[\rezprob(X_i(0))]$.
In view of \req{fast1111}, the optimal choice of the channel selection probabilities to maximize $\ex[\sum_{i=1}^{N}\rezprob(X_i(0))p_i^2]$ is to
let $p_1=1$ and $p_i=0$, $i=2,\ldots, N$.
\eproof

From \rthe{fast}, we know that
the minimum ETTR in the fast time-varying channel model is $\frac{1}{\ex[\rezprob(X_1(0))]}$.
This is achieved when each user selects the best channel  all the time.

\bsec{The Slow Time-Varying Channel Model}{slow_chmod}

In the previous section, we have shown that the optimal channel selection policy is to select the best channel all the time in the fast time-varying channel model. Thus, when all the $N$ channels are identically distributed, we can simply select channel $1$  all the time.
However, such a conclusion is no longer valid in the slow time-varying channel model where the state of each channel remains unchanged through the rendezvous process.

\bdefin{slow}{\bf (Slow Time-Varying Channel Model)}
For the slow time-varying channel model, the sequence of random vectors
$\{\boldsymbol{X}(t)=(X_1(t),X_2(t),\ldots, X_N(t)), t\ge 0\}$ do not change with respect to time, i.e.,
$$\boldsymbol{X}(t)=\boldsymbol{X}(0)$$
for all $t$. Moreover,
we assume that $\boldsymbol{X}(0)$ has the joint probability mass function $q(\boldsymbol{x})$, i.e.,
$$\pr(\boldsymbol{X}(0)=\boldsymbol{x})=q(\boldsymbol{x}).$$
\edefin

Suppose that $\boldsymbol{X}(0)=\boldsymbol{x}=(x_1,x_2,\ldots,x_N)$. Then under the slow time-varying channel model, the random variable ${T(\boldsymbol{p})}$ is geometrically distributed with parameter $\sum_{i=1}^{N}\rezprob(x_i)p_i^2$. Then the conditional expectation of ${T(\boldsymbol{p})}$ on $\boldsymbol{X}(0)=\boldsymbol{x}$ is
\bearn
	&& \ex[{T(\boldsymbol{p})}|\boldsymbol{X}(0)=\boldsymbol{x}] = 1/\sum_{i=1}^{N}\rezprob(x_i)p_i^2.
\eearn
This then leads to
\bear{ettr2222}
	&& \ex[{T(\boldsymbol{p})}]=\sum_{\boldsymbol{x}}\ex[{T(\boldsymbol{p})}|\boldsymbol{X}(0)=\boldsymbol{x}]P(\boldsymbol{X}(0)=\boldsymbol{x}) \nonumber \\
	&& =\sum_{\boldsymbol{x}}\frac{1}{\sum_{i=1}^{N}\rezprob(x_i)p_i^2}\ q(\boldsymbol{x}) \nonumber\\
&& =\ex \Big [\frac{1}{\sum_{i=1}^{N}\rezprob(X_i(0))p_i^2}\Big ].
\eear
For the slow time-varying channel model, the optimization problem  that minimizes the ETTR in \req{opti0011}
can be reformulated as follows:
\bear{opti1111}
&\min_{\boldsymbol{p}} \quad &\ex \Big [\frac{1}{\sum_{i=1}^{N}\rezprob(X_i(0))p_i^2} \Big ]\nonumber \\
&s.t. \quad &p_i \ge 0,\;i=1,2,\ldots,N, \nonumber \\
&&\sum_{i=1}^n p_i=1.
\eear

\bsubsec{Change of probability vectors}{change}

Minimizing the ETTR under the {\em slow} time-varying channel model is much more difficult than that in the {\em fast} time-varying channel model.
For this, we transform the minimization problem for ETTR in \req{opti0011} into an equivalent one by using the change of
probability vectors described in \rprop{change}.

\bprop{change}
Consider a probability vector ${\boldsymbol{p}}$.
For $i=1,2, \ldots, N$, let
\beq{change1111}
u_i =\frac{p_i^2}{\sum_{j=1}^N p_j^2},
\eeq
and  ${\boldsymbol{u}}=(u_1, u_2, \ldots, u_N)$.
Then ${\boldsymbol{u}}$ is also a probability vector. On the other hand,
for a probability vector $\boldsymbol{u}$, let
\beq{change2222}
p_i =\frac{\sqrt{u_i}}{\sum_{j=1}^N \sqrt{u_j}},
\eeq
for all $i=1,2 \ldots, N$ and ${\boldsymbol{p}}=(p_1,p_2, \ldots, p_N)$. Then ${\boldsymbol{p}}$ is also a probability vector.
\eprop

\bproof
Since $\boldsymbol{p}$ is a probability vector, we know that $\sum_{j=1}^N p_j^2 >0$.
In view of \req{change1111}, we know that $0 \le u_i \le 1$ and $\sum_{i=1}^N u_i=1$.
The argument for ${\boldsymbol{p}}$ defined in \req{change2222} to be a probability vector is similar.
\eproof

The mappings in \req{change1111} and \req{change2222} define a one-to-one transformation between the two probability vectors $\boldsymbol{p}$ and $\boldsymbol{u}$.
Moreover, from \req{change1111}, we know that
\beq{change3333}
\sum_{i=1}^N \sqrt{u_i} =\frac{1}{\sqrt{\sum_{j=1}^N p_j^2}}.
\eeq
With this in mind, we can rewrite the ETTR by using the probability vector $\boldsymbol{u}$ as follows:
\bear{ettr2222c}
	&& \ex[{T(\boldsymbol{p})}] =\ex \Big [\frac{1}{\sum_{i=1}^{N}\rezprob(X_i(0))p_i^2}\Big ] \nonumber \\
&&=\frac{1}{\sum_{j=1}^N p_j^2}\ex \Big [\frac{1}{\sum_{i=1}^{N}\rezprob(X_i(0))\frac{p_i^2}{\sum_{j=1}^N p_j^2}}\Big ]\nonumber\\
&&=(\sum_{i=1}^N \sqrt{u_i})^2 \ex \Big [\frac{1}{\sum_{i=1}^{N}\rezprob(X_i(0))u_i}\Big ]\nonumber\\
&&=
g({\boldsymbol{u}}) f({\boldsymbol{u}}),
\eear
where
\beq{ettr3377}
f({\boldsymbol{u}})=\ex \Big [\frac{1}{\sum_{i=1}^{N}\rezprob(X_i(0))u_i}\Big ]
\eeq
and
\beq{change4444}
g({\boldsymbol{u}})=(\sum_{i=1}^N \sqrt{u_i})^2.
\eeq
Now the problem to  minimize the ETTR in \req{opti0011}
can be reformulated as follows:
\bear{opti1111c}
&\min_{\boldsymbol{u}} \quad &g({\boldsymbol{u}}) f({\boldsymbol{u}})\nonumber \\
&s.t. \quad &u_i \ge 0,\;i=1,2,\ldots,N, \nonumber \\
&&\sum_{i=1}^N u_i=1.
\eear


\bsubsec{Majorization ordering and bounds for the ETTR}{lower}

In this section, we derive a lower bound and an upper bound for the ETTR when the channel states are {\em exchangeable random variables}.
Specifically, the $N$ random variables, $X_1(0),X_2(0)\ldots,X_N(0)$, are exchangeable random variables if their joint distribution
$q(\boldsymbol{x})$ is symmetric, i.e.,
$$q(\boldsymbol{x})=q(\boldsymbol{x^\pi})$$
 for any permutation $\pi=(\pi(1), \pi(2), \ldots, \pi(N))$ and $\boldsymbol{x^\pi}=(x_{\pi(1)},x_{\pi(2)},\ldots,x_{\pi(N)})$.
 Clearly, if the $N$ random variables, $X_1(0),X_2(0)\ldots, X_N(0)$, are independent and identically distributed, then
they are also exchangeable random variables.

Our approach is based on  the theory of majorization ordering in \cite{MarOl}.

\bdefin{major}
A vector ${\bf x}=(x_1, x_2, \ldots, x_N)$ is
{\em majorized} by another vector ${\bf y}=(y_1,y_2, \ldots, y_N)$ (${\bf x} \prec {\bf y}$) if
(i) $\sum_{i=1}^j{x_{[i]}} \le
\sum_{i=1}^j{y_{[i]}}$, $j=1, \ldots , N-1$ and (ii)
$\sum_{i=1}^N{x_{[i]}} =
\sum_{i=1}^N{y_{[i]}}$, where $x_{[i]}$ ($y_{[i]}$) is the $i$-th
largest component in $\bf x$ ($\bf y$).
\edefin

Intuitively, majorization ordering is a partial ordering that indicates whether a vector is more ``balanced'' than another.
For example,
\bearn
&&(\frac{1}{N}, \ldots, \frac{1}{N}) \prec (\frac{1}{N-1}, \ldots, \frac{1}{N-1},0) \\
 && \prec (\frac{1}{2}, \frac{1}{2}, 0, \ldots, 0) \prec ({1}, 0, 0, \ldots, 0).
\eearn
Since $(\frac{1}{N}, \ldots, \frac{1}{N}) \prec {\bf x} \prec ({1}, 0, 0, \ldots, 0)$, the vector $(\frac{1}{N}, \ldots, \frac{1}{N})$ is the most balanced vector and the vector $({1}, 0, 0, \ldots, 0)$ is the most unbalanced vector.
Majorization ordering  has several equivalent characterizations (see e.g., \cite{MarOl}, pp. 11 for a summary of some majorization equivalents). In the following proposition, we list some of them that will be used in this paper.

\bprop{majorprop}
For majorizations, the following conditions are equivalent:
\begin{description}
\item[(i)] ${\bf x} \prec {\bf y}$.
\item[(ii)] $h({\bf x}) \le h({\bf y})$ for all symmetric convex functions $h$ on ${\cal R}^N$.
\item[(iii)] $\sum_i h(x_i) \le \sum_i h(y_i)$ for all convex functions $h$ on ${\cal R}$.
\end{description}
\eprop

\bdefin{majorfun}
A function $h:{\cal R}^N \mapsto {\cal R}$ is said to be Schur convex (resp. concave) if
$$
{\bf x} \prec {\bf y} \Rightarrow h({\bf x}) \le h({\bf y}) \quad (\mbox{resp.}\; h({\bf x}) \ge h({\bf y})).
$$
\edefin

It follows immediately from \rprop{majorprop} that
symmetric convex functions are Schur convex and
separable convex (resp. concave) functions are also Schur convex (resp. concave).
Clearly,
$\sum_{i=1}^N \sqrt{u_i}$ is the sum of separable concave functions and thus a Schur concave function.
This then implies that $g({\boldsymbol{u}})$ is Schur concave and thus
\beq{lower0058}
N=g(\frac{1}{N}, \ldots, \frac{1}{N}) \ge g({\boldsymbol{u}}) \ge g({1}, 0, 0, \ldots, 0)=1.
 \eeq

In the following lemma, we show that the function $f({\boldsymbol{u}})$ is Schur convex and use that to derive bounds for the ETTR.

\blem{lower}
Suppose that the $N$ random variables, $X_1(0),X_2(0)\ldots,X_N(0)$, are exchangeable random variables.
\begin{description}
\item[(i)]
The function $f(\boldsymbol{u})$ in \req{ettr3377} is symmetric and convex in $\boldsymbol{u}$ and thus Schur convex in $\boldsymbol{u}$.
\item[(ii)] The ETTR has the following lower bound and upper bound:
\bear{ettr3355}
&&\ex\Big [\frac{N}{\sum_{i=1}^N\rezprob(X_i(0))} \Big] \le 	 \ex[{T(\boldsymbol{p})}] \nonumber\\
&&\le  N \ex[\frac{1}{\rezprob({X}_1(0))} ].
\eear
\end{description}
\elem

\bproof
(i)
Let $\pi^{-1}$ be the inverse permutation of $\pi$ and $\boldsymbol{u^\pi}=(u_{\pi(1)},u_{\pi(2)},\ldots,u_{\pi(N)})$.
For a vector ${\bf x}$ and a permeation $\pi$, let
$\rezprob(\boldsymbol{x^\pi})=(\rezprob(x_{\pi(1)}),\rezprob(x_{\pi(2)})),\ldots,\rezprob(x_{\pi(N)}))$.
Since the $N$ random variables, $X_1(0),X_2(0)\ldots,X_N(0)$, are exchangeable random variables, we know that for any permutation $\pi$ of $(1,2,\ldots, N)$
\bear{ettr4422}
&&f(\boldsymbol{u})=\ex[\frac{1}{\rezprob(\boldsymbol{X}(0)) \cdot \boldsymbol{u}}]\nonumber\\
&&=\sum_{\boldsymbol{x}}\frac{1}{\rezprob(\boldsymbol{x}) \cdot \boldsymbol{u}} q(\boldsymbol{x}) \nonumber\\
&&=\sum_{\boldsymbol{y}}\frac{1}{\rezprob(\boldsymbol{y^{\pi^{-1}}}) \cdot \boldsymbol{u}} q(\boldsymbol{y^{\pi^{-1}}}) \quad \mbox{(change of variables)}   \nonumber\\
&&=\sum_{\boldsymbol{y}}\frac{1}{\rezprob(\boldsymbol{y^{\pi^{-1}}}) \cdot \boldsymbol{u}} q(\boldsymbol{y}) \quad \mbox{(exchangeability)}.
\eear
Note that
\bearn
\rezprob(\boldsymbol{y^{\pi^{-1}}}) \cdot \boldsymbol{u}
&=&\sum_{i=1}^N \rezprob(y_{\pi^{-1}(i)}) \;u_i \\
&=&\sum_{j=1}^N \rezprob(y_{j})\; u_{\pi(j)}
=\rezprob(\boldsymbol{y}) \cdot \boldsymbol{u^\pi}.
\eearn
In conjunction with \req{ettr4422},
\bear{ettr4455}
f(\boldsymbol{u})
&=&\sum_{\boldsymbol{y}}\frac{1}{\rezprob(\boldsymbol{y) \cdot \boldsymbol{u^{\pi}}}} q(\boldsymbol{y}) \nonumber\\
&=&\ex[\frac{1}{\rezprob(\boldsymbol{X}(0)) \cdot \boldsymbol{u^\pi}}].
\eear
As \req{ettr4455} holds for any permutation $\pi$, it then follows that
\beq{ettr4466}
f(\boldsymbol{u})=\frac{1}{N!}\sum_{\pi \in \Pi}\ex[\frac{1}{\rezprob(\boldsymbol{X}(0)) \cdot \boldsymbol{u^\pi}}],
\eeq
where $\Pi$ is the set of $N!$ permutations of $(1,2, \ldots, N)$.
Clearly, the right-hand side of \req{ettr4466} is symmetric  in $\boldsymbol{u}$ as the sum is over all the permutations $\pi \in \Pi$.
Thus,  $f(\boldsymbol{u})$ is symmetric  in $\boldsymbol{u}$.

To prove $f(\boldsymbol{u})$ is convex, we need to show that
\beq{lower2255}
f(\alpha \boldsymbol{u^\prime} + (1-\alpha) \boldsymbol{u^{\prime\prime}})\leq \alpha f(\boldsymbol{u^{\prime}}) + (1-\alpha) f(\boldsymbol{u^{\prime\prime}}), \eeq
for $0 \leq \alpha \leq 1$ and any two $N$-vectors $\boldsymbol{u^\prime}$ and $\boldsymbol{u^{\prime\prime}}$.
To see this, note that the function $h(x)=1/x$ is convex in $x$. Thus,
\bearn
	&& \frac{1}{\rezprob(\boldsymbol{X}(0)) \cdot (\alpha \boldsymbol{u^\prime}+(1-\alpha)\boldsymbol{u^{\prime\prime}})}\\
 &&= \frac{1}{\alpha \rezprob(\boldsymbol{X}(0))\cdot \boldsymbol{u^\prime}+(1-\alpha)\rezprob(\boldsymbol{X}(0))\cdot \boldsymbol{u^{\prime\prime}}} \\
	&& \leq \alpha \frac{1}{\rezprob(\boldsymbol{X}(0)) \cdot \boldsymbol{u^\prime}}+(1-\alpha)\frac{1}{\rezprob(\boldsymbol{X}(0)) \cdot \boldsymbol{u^{\prime\prime}}}.
\eearn
Taking expectations on both sides of the above inequality yields \req{lower2255}.

(ii)
As shown in \rlem{lower}(i), the function $f({\boldsymbol{u}})$ is
symmetric and convex and thus a Schur convex function. Thus, we have from the majorization ordering that
\bear{lower0056}
&&\ex\Big [\frac{N}{\sum_{i=1}^N\rezprob(X_i(0))} \Big]=f(\frac{1}{N}, \ldots, \frac{1}{N}) \nonumber \\
 &&\le f({\boldsymbol{u}}) \le f({1}, 0, 0, \ldots, 0)=\ex[\frac{1}{\rezprob({X}_1(0))} ].
 \eear
From \req{lower0056} and \req{lower0058}, we then have
$$\ex\Big [\frac{N}{\sum_{i=1}^N\rezprob(X_i(0))} \Big] \le g({\boldsymbol{u}})f({\boldsymbol{u}}) \le N \ex[\frac{1}{\rezprob({X}_1(0))} ].$$
The upper bound and the lower bound for ETTR in \req{ettr3355} then follows from the representation of the ETTR in \req{ettr2222c},

\eproof

\bsubsec{Approximation solutions}{solutions}

In this section, we propose approximation solutions for the ETTR minimization problem in \req{opti1111}.
Note that an $N$-approximation solution of a minimization problem is referred to as a solution that is not greater than the $N$ times of the
optimal solution.

\bthe{approx}
Suppose that the $N$ random variables, $X_1(0),X_2(0)\ldots,X_N(0)$, are exchangeable random variables, i.e.,
$q(\boldsymbol{x})=q(\boldsymbol{x^\pi})$ for any permutation $\pi$,
\begin{description}
\item[(i)] The uniform selection policy that uses $p_i=1/N$ for all $i=1,2, \ldots, N$,
is an $N$-approximation solution for the ETTR minimization problem in \req{opti1111}.
\item[(ii)] Let
\beq{approx0022}
M=\frac{\ex[\frac{1}{\rezprob({X}_1(0))} ]}{\frac{1}{\ex[{\rezprob({X}_1(0))} ]}}.
\eeq
The single selection policy that uses $p_1=1$ and $p_i=0$, $i=2,3, \ldots, N$, is an
$M$-approximation solution for the ETTR minimization problem in \req{opti1111}.
\item[(iii)] The policy that uses the better one between the single selection policy and the uniform selection policy is
a $\min[M,N]$-approximation solution for the ETTR minimization problem in \req{opti1111}.
\end{description}
\ethe

\bproof
(i)
Note that if we use the uniform selection policy $p_i=1/N$ for all $i=1,2, \ldots, N$ in \req{ettr2222}, then the corresponding ETTR  is
\beq{approx1111}
\ex[\frac{N^2}{\sum_{i=1}^{N}\rezprob(X_i(0))}].
\eeq
It then follows from the lower bound in \rlem{lower}(ii) that the ETTR in \req{approx1111} is not greater than
the $N$ times of the
optimal solution for the ETTR minimization problem in \req{opti1111}.

(ii) Since $h(x)=1/x$ is a convex function in $x$, it then follows from Jensen's inequality (see e.g., the book \cite{Ross1996}) and the assumption of the $N$ exchangeable random variables that
\bear{approx1100}
&&\ex[\frac{N}{\sum_{i=1}^{N}\rezprob(X_i(0))}] \ge \frac{N}{\ex[\sum_{i=1}^{N}\rezprob(X_i(0))]}\nonumber \\
&&={\frac{1}{\ex[{\rezprob({X}_1(0))} ]}}.
\eear
On the other hand, we have from \req{ettr2222} that the ETTR of the single selection policy is
\beq{approx1122}
{\ex[\frac{1}{\rezprob({X}_1(0))} ]}=M {\frac{1}{\ex[{\rezprob({X}_1(0))} ]}}.
\eeq
It then follows from the lower bound in \rlem{lower}(ii) and \req{approx1100} that the ETTR in \req{approx1122}
is not greater than
the $M$ times of the
optimal solution for the ETTR minimization problem in \req{opti1111}.

(iii) This is a direct consequence of (ii) and (iii).
\eproof

The approximation ratio for \rthe{approx}(i) cannot be further improved. To see this, consider the case with $\rezprob(x)=1$ for all $x$, i.e.,
with probability 1 the two users will have a successful rendezvous when they hop on the same channel. In this case, the optimal policy is
the single selection policy, i.e., $p_1=1$ and $p_i=0$, $i=2, \ldots, N$,  with the TTR=1. However, the ETTR for the uniform selection policy
is $N$ and the approximation ratio for this case is $N$.


\bsec{A slow time-varying channel model with two states}{two}

The slow time-varying channel model considered in the previous section is too general to further improve the approximation results.
In this section, we consider a specific slow time-varying channel model with two states, i.e., state 0 (bad state) and state 1 (good state).
For the two-state model, we assume that the states of these $N$ channels are {\em independent and identically distributed}. The probability that a channel is in the good (resp. bad) state is $\rho$ (resp. $1-\rho$) for some $0 \le \rho \le 1$.
As such,  we have the following joint distribution for the channel states
\beq{twojoint111}
q(x_1,x_2, \ldots, x_N)=\prod_{i=1}^N \rho^{x_i} (1-\rho)^{1-x_i},
\eeq
where $\rho$ (resp. $(1-\rho)$) is the probability of being in state 1 (resp. 0), and $x_i$ (with the value being 0 or 1) is the state of channel $i$.

\bsubsec{Two channels}{two channels}

We start from the simplest case with two channels, i.e., $N=2$.
In this case, the joint probability mass function $q(x_1,x_2)$ can be characterized as follows:
\bearn
	&& q(0,0)=(1-\rho)^2,   \quad \ q(1,0)=\rho(1-\rho), \\
	&& q(0,1)=\rho(1-\rho), \quad \ q(1,1)=\rho^2.	
\eearn
For the blind rendezvous policy that each user selects channel 1 (resp. 2) with probability $p_1$ (resp. $p_2$), we have
\bear{ettr1111}
	 \ex[{T(\boldsymbol{p})}] &=& \frac{1}{\rezprob(0)p_1^2+\rezprob(0)p_2^2}(1-\rho)^2 \nonumber \\
&& +\frac{1}{\rezprob(1)p_1^2+\rezprob(0)p_2^2}\rho(1-\rho) \nonumber \\
	&&  +\frac{1}{\rezprob(0)p_1^2+\rezprob(1)p_2^2}\rho(1-\rho) \nonumber\\
&&+\frac{1}{\rezprob(1)p_1^2+\rezprob(1)p_2^2}\rho^2.
\eear

\begin{figure}[ht]
  \centering
  \begin{subfigure}[b]{0.45\linewidth}
    \includegraphics[width=\linewidth]{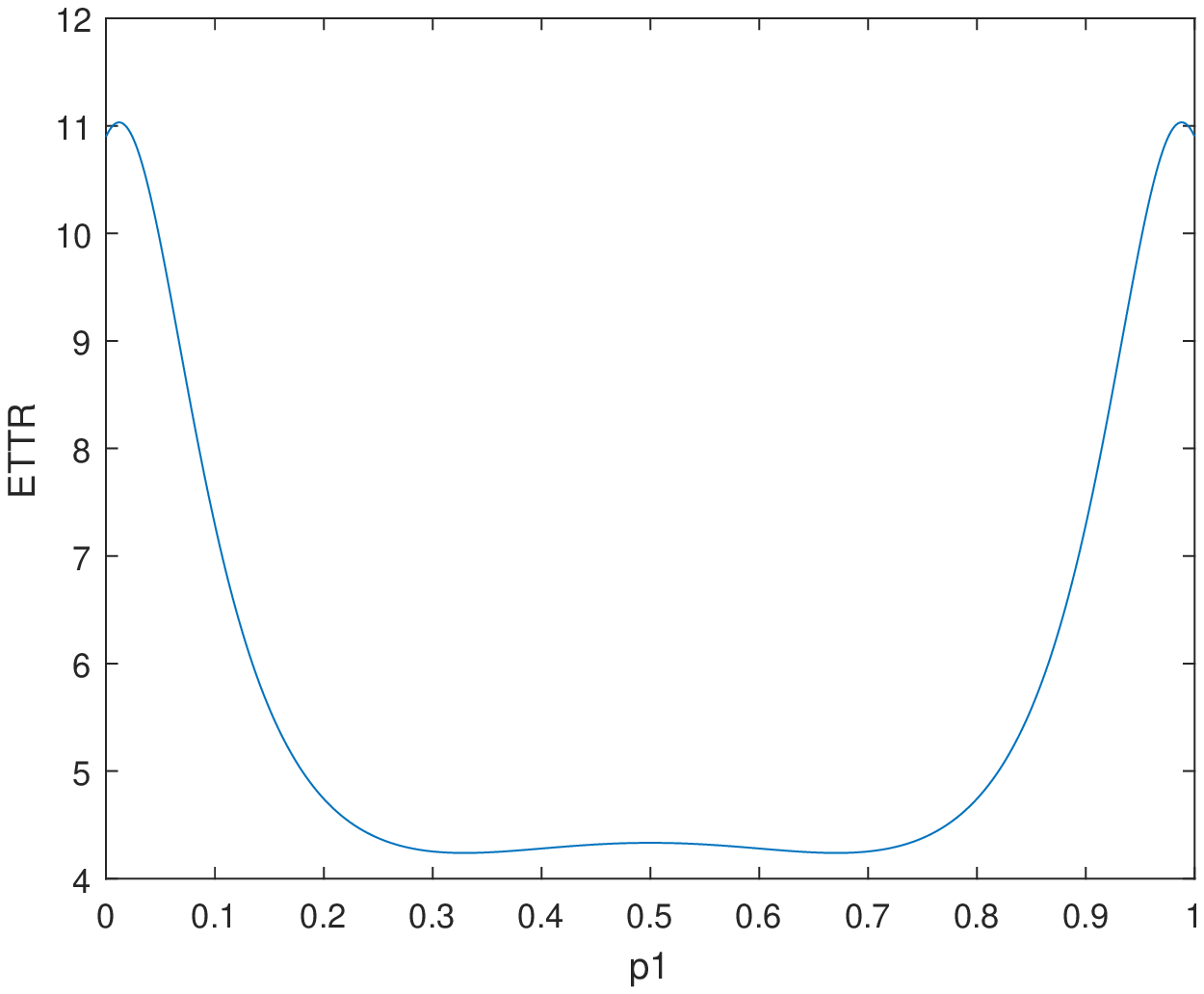}
     \caption{$\rezprob(0)$=0.01}
  \end{subfigure}
  \begin{subfigure}[b]{0.45\linewidth}
    \includegraphics[width=\linewidth]{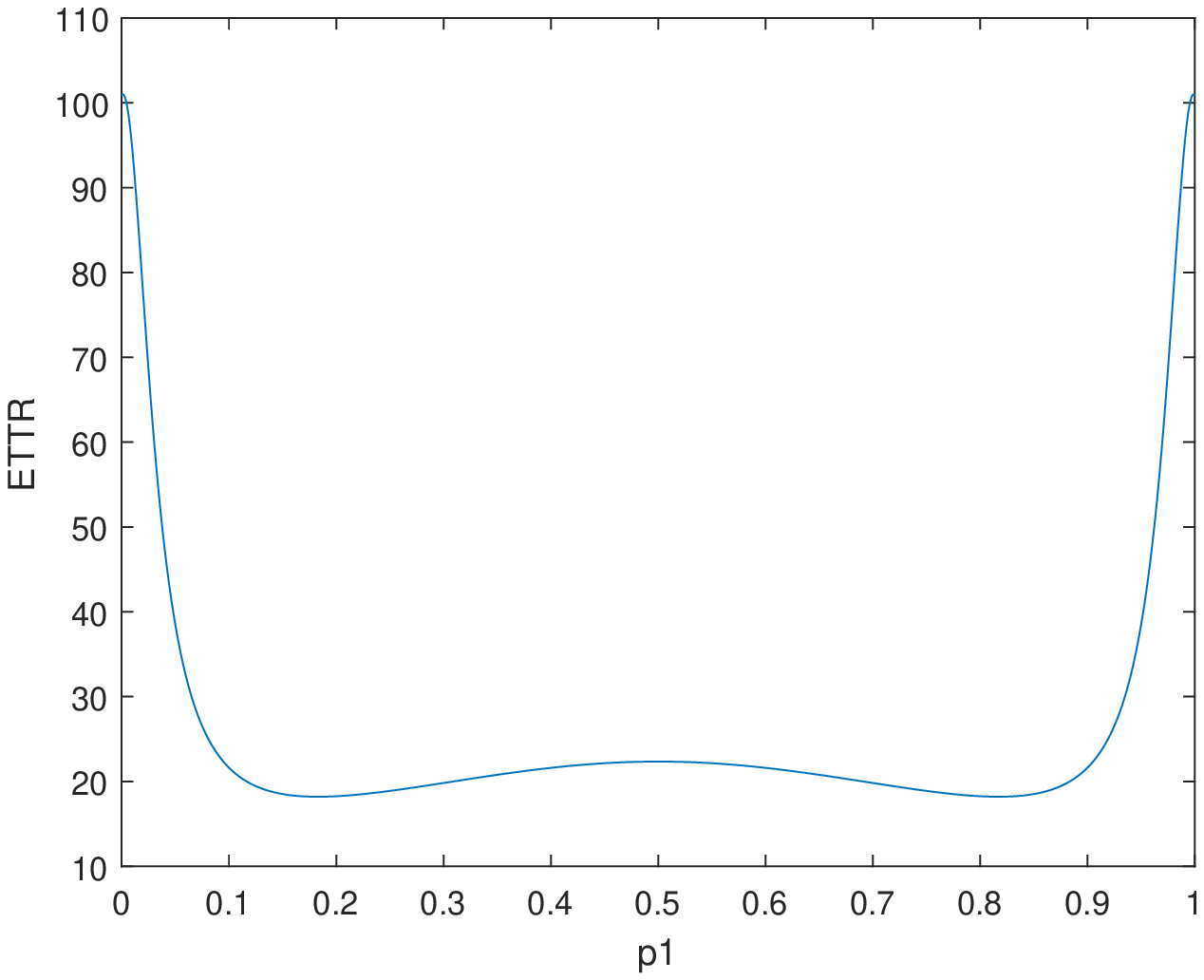}
    \caption{$\rezprob(0)$=0.001}
  \end{subfigure}
  \caption{The ETTR (as a function of $p_1$) with $\rho=0.9$, $\rezprob(1)=1$, and $\rezprob(0)=0.01, 0.001$, respectively.}
  \label{fig:ex1_indep_2_1}
\end{figure}

\begin{figure}[ht]
  \centering
  \begin{subfigure}[b]{0.45\linewidth}
    \includegraphics[width=\linewidth]{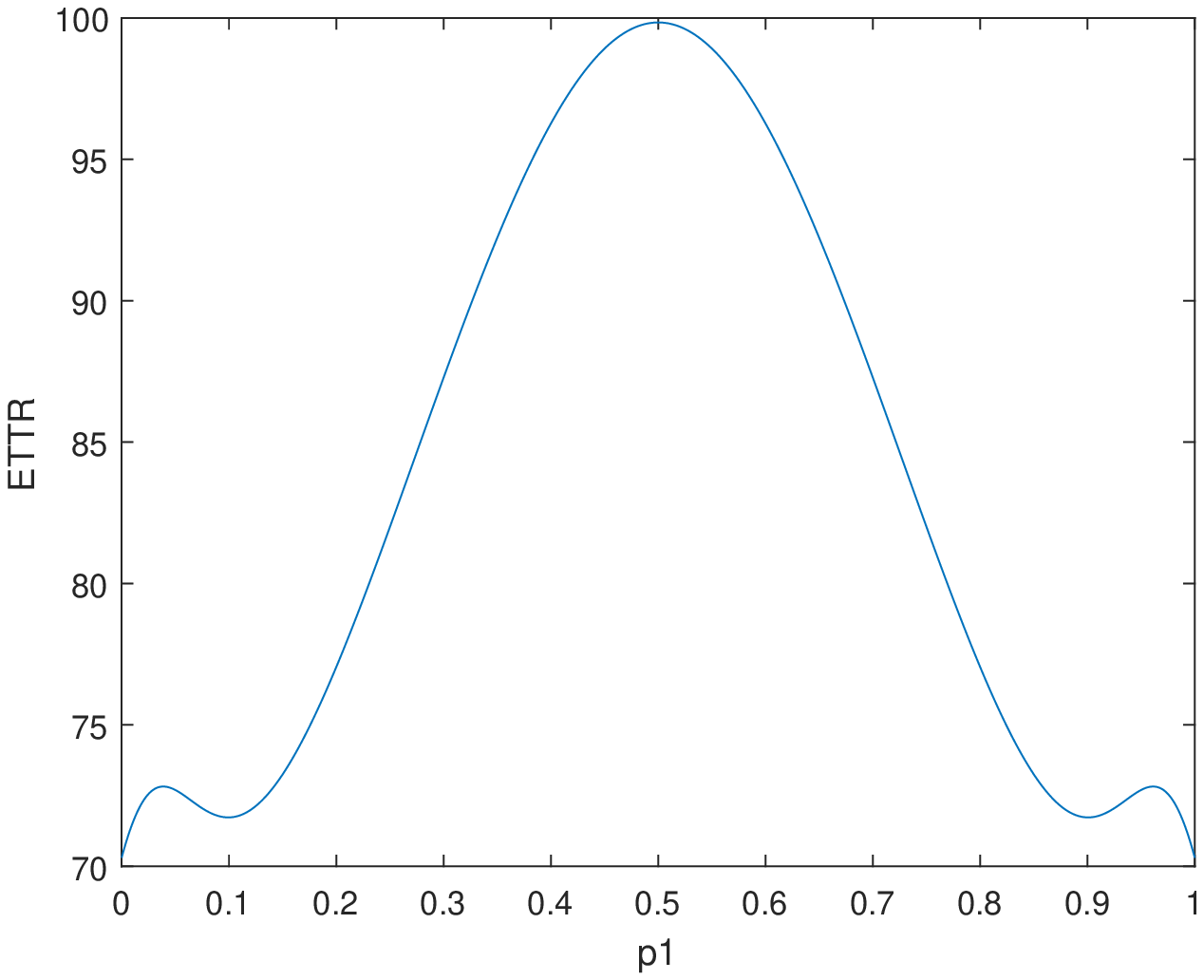}
     \caption{$\rezprob(0)$=0.01}
  \end{subfigure}
  \begin{subfigure}[b]{0.45\linewidth}
    \includegraphics[width=\linewidth]{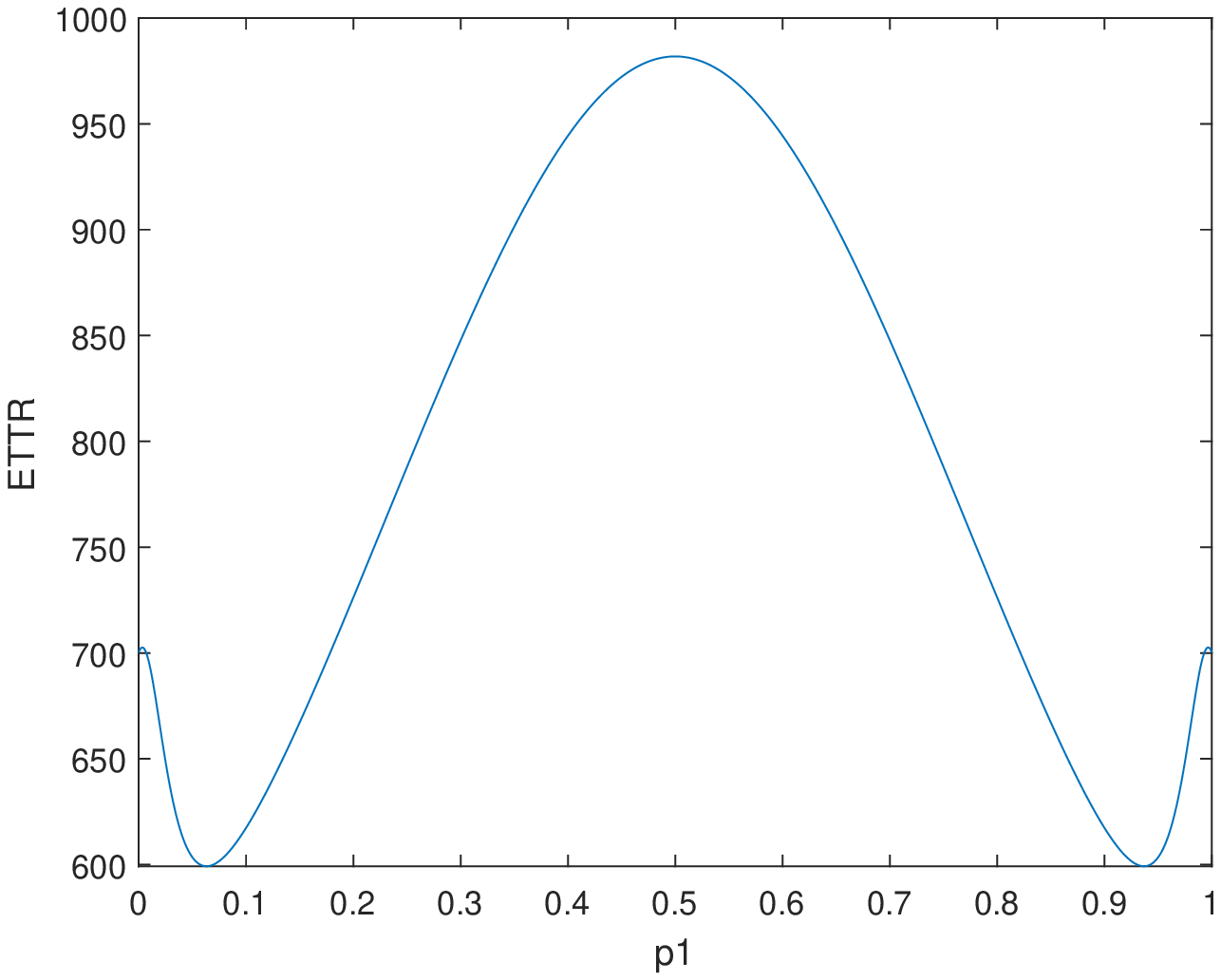}
    \caption{$\rezprob(0)$=0.001}
  \end{subfigure}
  \caption{The ETTR (as a function of $p_1$) with $\rho=0.3$, $\rezprob(1)=1$, and $\rezprob(0)=0.01, 0.001$, respectively.}
  \label{fig:ex1_indep_2_2}
\end{figure}

To gain some insights of \req{ettr1111},  we show in \rfig{ex1_indep_2_1} the numerical results for the ETTR when $\rho=0.9$, $\rezprob(1)=1$, and $\rezprob(0)$ is selected from $0.01$ and $0.001$, respectively. As shown in \rfig{ex1_indep_2_1}, the optimal channel selection policy that minimizes the ETTR is not the single selection policy that sets $p_1=1$ (or $p_1=0$).
Such a result can also be found for $\rho=0.3$ in \rfig{ex1_indep_2_2}(b).
To see the intuition behind this, note that if both users set $p_1=1$ and hop to channel 1 all the time, it is possible that channel 1 is in the bad state with a very low rendezvous probability. This then leads to a very large ETTR. As such, it is preferable to having a nonzero probability to hop on the other channel.

\begin{figure}[ht]
\centering
	\includegraphics[width=0.5\textwidth]{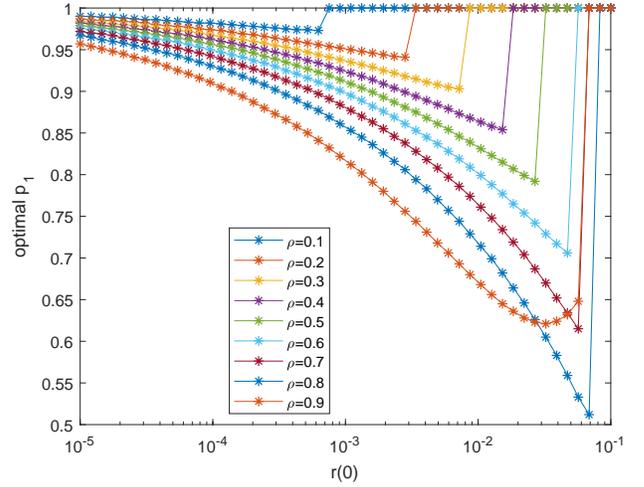}
	\caption{The optimal $p_1$ for $\rezprob(0)$ ranging from $0.1$ to $0.00001$ with $\rezprob(1)=1$.}
\label{fig:optimalp1}
\end{figure}

In view of \rfig{ex1_indep_2_1} and \rfig{ex1_indep_2_2}, the optimal $p_1$ clearly depends on how bad the bad state is and it is thus a function of the rendezvous probability for state 0, i.e., $\rezprob(0)$.
In \rfig{optimalp1}, we show the optimal $p_1$ (obtained from our numerical results) for $\rezprob(0)$ ranging from $0.1$ to $0.00001$ for various $\rho$'s.
It is interesting to see that the optimal $p_1$ is very close to 1 when $\rezprob(0)$ is either very small or very close to $\rezprob(1)$.
Such asymptotic results will be formally proved in the next section.


\bsubsec{An asymptotic $(1+\epsilon)$-approximation solution}{epsilon}

In this section, we show an asymptotic $(1+\epsilon)$-approximation solution for the two-state slow time-varying channel model with $N \ge 2$ channels.

Note from the joint distribution in \req{twojoint111} that
\bear{num1111}
&&f({\boldsymbol{u}})=\ex[\frac{1}{\sum_{i=1}^N r(X_i(0))u_i}]\nonumber \\
&&=\sum_{x_1=0}^1 \cdots \sum_{x_N=0}^1 \frac{1}{\sum_{i=1}^N r(x_i)u_i}q(x_1,x_2, \ldots, x_N)\nonumber \\
&&=\sum_{x_1=0}^1 \cdots \sum_{x_N=0}^1 \frac{1}{\sum_{i=1}^N r(x_i)u_i}\prod_{i=1}^N \rho^{x_i} (1-\rho)^{1-x_i}.\nonumber \\
\eear
Thus, the computational complexity for $f({\boldsymbol{u}})$ is $O(2^N)$. So is the computational complexity for
$\ex[{T(\boldsymbol{p})}]$. As it is very costly to compute $\ex[{T(\boldsymbol{p})}]$ directly for large $N$, in the following lemma we derive
two asymptotic results  to gain some insights of the ETTR.

\blem{two}
Consider the two-state slow time-varying channel model with the joint distribution in \req{twojoint111}.
Suppose that the three parameters $r(1), \rho$ and $N$ are fixed constants.
\begin{description}
\item[(i)] For any probability vector $\boldsymbol{u}$,
\beq{two0011a}\lim_{r(0) \to r(1)} r(0)f({\boldsymbol{u}})= 1.
\eeq
\item[(ii)] For any probability vector $\boldsymbol{u}$ with $u_i>0$, $i=1,2, \ldots, N$,
\beq{two0011b}
\lim_{r(0) \to 0} r(0)f({\boldsymbol{u}})/(1-\rho)^N =1.
\eeq
\end{description}
As a result, for any probability vector $\boldsymbol{p}$,
\bear{two0022}
&&\lim_{r(0) \to r(1)} r(0)\ex[T(\boldsymbol{p})]\ge 1, \label{eq:two0022a} \\
&&\lim_{r(0) \to 0} r(0)\ex[T(\boldsymbol{p})]/(1-\rho)^N\ge 1.\label{eq:two0022b}
\eear
\elem

\bproof
Since $r(0) \le r(x_i) \le r(1)$, it follows from \req{num1111} that
$$\frac{1}{r(1)} \le f({\boldsymbol{u}}) \le \frac{1}{r(0)}.$$
Taking the limit then completes the argument for \req{two0011a}.

Now we prove \req{two0011b} under the condition that $u_i >0$ for all $i=1,2,\ldots, N$.
Since $r(1), \rho$ and $N$ are fixed constants,
we have for any ${\bf x}=(x_1,x_2,\ldots, x_N)\ne (0,0,\ldots,0)$ that
\beq{two9988}
\lim_{r(0) \to 0}\frac{\frac{1}{\sum_{i=1}^N r(x_i)u_i}\prod_{i=1}^N \rho^{x_i} (1-\rho)^{1-x_i}}{\frac{1}{\sum_{i=1}^N r(0)u_i}(1-\rho)^N}
=0.
\eeq
Since $\sum_{i=1}^N u_i=1$, it then follows from \req{num1111} and \req{two9988} that
$$\lim_{r(0) \to 0}\frac{f({\boldsymbol{u}})}{\frac{1}{ r(0)}(1-\rho)^N}
=0.$$

From \req{lower0058}, we also know that
$g({\boldsymbol{u}}) \ge 1$.
Since $\ex[T(\boldsymbol{p})] =g({\boldsymbol{u}})f({\boldsymbol{u}})$ in \req{ettr2222c}, we have from \req{two0011a} that
\bear{two2233}
&&\lim_{r(0) \to r(1)}  r(0)\ex[T(\boldsymbol{p})] \nonumber \\
&&=\lim_{r(0) \to r(1)}  r(0)g({\boldsymbol{u}})f({\boldsymbol{u}})\ge 1.
\eear


Again, using $g({\boldsymbol{u}}) \ge 1$ and \req{two0011b} yields
\bear{two2244}
&&\lim_{r(0) \to 0}  r(0)\ex[T(\boldsymbol{p})]/(1-\rho)^N \nonumber \\
&&=\lim_{r(0) \to 0}  r(0)g({\boldsymbol{u}})f({\boldsymbol{u}})/(1-\rho)^N \ge 1
\eear
for all ${\boldsymbol{u}}$ with $u_i >0$, $i=1,2, \ldots, N$.
Now consider the case that $u_N=0$ and $u_i>0$ for all $i=1,2,\ldots, N-1$. Then this is equivalent to the case that there are only $N-1$ channels (as channel $N$ is not used). As a direct sequence of \req{two2244}, we have for this case that
\bear{two2255}
&&\lim_{r(0) \to 0}  r(0)\ex[T(\boldsymbol{p})]/(1-\rho)^{N-1}
\ge 1.
\eear
Since $(1-\rho) \le 1$, the inequality in \req{two2255} also implies the inequality in \req{two0022b}.
Repeating the same argument shows that the inequality in \req{two0022b} holds for  any probability vector ${\boldsymbol{p}}$.
\eproof

Now we use the asymptotic results in \rlem{two} to derive an asymptotic $(1+\epsilon)$-approximation solution.

\bthe{twomain}
Consider the two-state slow time-varying channel model with the joint distribution in \req{twojoint111}.
For any $0< \epsilon \le 3$, let $\delta=(\frac{\epsilon}{3(N-1)})^2$ and consider the probability vector
\beq{two6666}
{\boldsymbol{u}}^*=(u_1^*, u_2^*, \ldots, u_N^*)=(1-(N-1)\delta, \delta, \ldots, \delta).
\eeq
Let ${\boldsymbol{p}}^*=(p_1^*, p_2^*, \ldots, p_N^*)$ with
\beq{change2222m}
p_i^* =\frac{\sqrt{u_i^*}}{\sum_{j=1}^N \sqrt{u_j^*}}.
\eeq
Then ${\boldsymbol{p}}^*$ is an asymptotic $(1+\epsilon)$-approximation solution for the ETTR minimization problem in \req{opti0011}
in the asymptotic regime when $r(0) \to r(1)$ or $r(0) \to 0$. Specifically,
for any probability vector ${\boldsymbol{p}}$,
\bear{two0022m}
&&\lim_{r(0) \to r(1)} \frac{\ex[T(\boldsymbol{p^*})]}{\ex[T(\boldsymbol{p})]} \le 1+\epsilon, \label{eq:two0022ma} \\
&&\lim_{r(0) \to 0} \frac{\ex[T(\boldsymbol{p^*})]}{\ex[T(\boldsymbol{p})]} \le 1+\epsilon.\label{eq:two0022mb}
\eear
\ethe

\bproof
Since $\ex[T(\boldsymbol{p})]=g({\boldsymbol{u}})f({\boldsymbol{u}})$, it follows that
\beq{two7777}
 \frac{\ex[T(\boldsymbol{p^*})]}{\ex[T(\boldsymbol{p})]} =\frac{g({\boldsymbol{u}}^*)}{g({\boldsymbol{u}})} \frac{f({\boldsymbol{u}}^*)}{f({\boldsymbol{u}})}.
 \eeq
 From \req{two0011a}, we then have
\beq{two7788}
\lim_{r(0) \to r(1)} \frac{f({\boldsymbol{u}}^*)}{f({\boldsymbol{u}})}=1.
\eeq
On the other hand, note from $0< \epsilon \le 3$ that
$$0< \delta \le 1/(N-1)^2$$
 and
\beq{two8811}
(N-1)^2\delta=\frac{\epsilon^2}{9} \le \frac{\epsilon}{3}.
\eeq
Using \req{change4444} and \req{two8811} yields
\bearn
&&g({\boldsymbol{u}}^*)=(\sum_{i=1}^N \sqrt{u_i^*})^2 \\
&&=(\sqrt{1-(N-1)\delta}+(N-1)\sqrt{\delta})^2 \\
&&\le 1-(N-1)\delta+(N-1)^2\delta +2(N-1) \sqrt{\delta} \\
&&\le 1+ \frac{\epsilon}{3}+\frac{2\epsilon}{3}=1+\epsilon.
\eearn
As $g({\boldsymbol{u}}) \ge 1$, we then have
\beq{two7799}
\frac{g({\boldsymbol{u}}^*)}{g({\boldsymbol{u}})} \le 1+\epsilon.
\eeq
The result in \req{two0022ma} then follows from \req{two7777}, \req{two7788} and \req{two7799}.

The argument for \req{two0022mb} is similar (by using \req{two0011b} and \req{two0022b}).
\eproof

In Table \ref{table:p_dis}, we use the grid search to find the optimal channel selection probability vector for the two-state model with three channels, i.e., $N=3$.
The grid search is conducted by using the resolution size 0.001, i.e.,
we compute the ETTR for all the channel selection probability vectors $\boldsymbol{p}$ with $p_i$'s being integer multiples of 0.001, and then select the minimum ETTR. As shown in Table \ref{table:p_dis}, the optimal $\boldsymbol{p}$ is very close to the single selection policy in most settings, i.e., the users will select a single channel with a very high probability $p_1$, and then select the rest of the channels with probability $(1-p_1)/(N-1)$.  This is consistent with the asymptotic $(1+\epsilon)$-approximation solution in \rthe{twomain}.
However, we also note from Table \ref{table:p_dis} (e) for the case $\rezprob(0)=10^{-3} \sim 10^{-2}$ and $\rho=0.9$, the optimal channel selection probability vector $\boldsymbol{p}$ is not in the form of the asymptotic optimal policy in \rthe{twomain}.

\begin{table}
\makebox[0.87\linewidth][c]{
\parbox{.5\linewidth}{
\centering
\begin{subtable}[c]{.3\textwidth}
\centering
\resizebox{0.8\textwidth}{10mm}{
\begin{tabular}{|l|l|l|l|}
\hline
\textbf{r(0)}                       & \textbf{p1} & \textbf{p2} & \textbf{p3} \\ \hline
\textbf{10\textasciicircum{}(-5)}   & 0.965       & 0.018       & 0.018       \\ \hline
\textbf{10\textasciicircum{}(-4.8)} & 0.965       & 0.018       & 0.018       \\ \hline
\textbf{10\textasciicircum{}(-4.6)} & 0.965       & 0.018       & 0.018       \\ \hline
\textbf{10\textasciicircum{}(-4.4)} & 0.965       & 0.018       & 0.018       \\ \hline
\textbf{10\textasciicircum{}(-4.2)} & 0.965       & 0.018       & 0.018       \\ \hline
\textbf{10\textasciicircum{}(-4)}   & 0.965       & 0.018       & 0.018       \\ \hline
\end{tabular}}
\caption{$\rezprob(0)=10^{-5} \sim 10^{-4}$}
\end{subtable}

\begin{subtable}[c]{.3\textwidth}
\centering
\resizebox{0.8\textwidth}{10mm}{
\begin{tabular}{|l|l|l|l|}
\hline
\textbf{r(0)}                       & \textbf{p1} & \textbf{p2} & \textbf{p3} \\ \hline
\textbf{10\textasciicircum{}(-3)}   & 1.00        & 0.00        & 0.00        \\ \hline
\textbf{10\textasciicircum{}(-2.8)} & 1.00        & 0.00        & 0.00        \\ \hline
\textbf{10\textasciicircum{}(-2.6)} & 1.00        & 0.00        & 0.00        \\ \hline
\textbf{10\textasciicircum{}(-2.4)} & 1.00        & 0.00        & 0.00        \\ \hline
\textbf{10\textasciicircum{}(-2.2)} & 1.00        & 0.00        & 0.00        \\ \hline
\textbf{10\textasciicircum{}(-2)}   & 1.00        & 0.00        & 0.00        \\ \hline
\end{tabular}}
\caption{$\rezprob(0)=10^{-3} \sim 10^{-2}$}
\end{subtable}

\begin{subtable}[c]{.3\textwidth}
\centering
\resizebox{0.8\textwidth}{10mm}{
\begin{tabular}{|l|l|l|l|}
\hline
\textbf{r(0)}                        & \textbf{p1} & \textbf{p2} & \textbf{p3} \\ \hline
\textbf{10\textasciicircum{}(-0.1)}  & 1.00        & 0.00        & 0.00        \\ \hline
\textbf{10\textasciicircum{}(-0.08)} & 1.00        & 0.00        & 0.00        \\ \hline
\textbf{10\textasciicircum{}(-0.06)} & 1.00        & 0.00        & 0.00        \\ \hline
\textbf{10\textasciicircum{}(-0.04)} & 1.00        & 0.00        & 0.00        \\ \hline
\textbf{10\textasciicircum{}(-0.02)} & 1.00        & 0.00        & 0.00        \\ \hline
\textbf{10\textasciicircum{}(0)}     & 1.00        & 0.00        & 0.00        \\ \hline
\end{tabular}}
\caption{$\rezprob(0)=10^{-0.1} \sim 1$}
\end{subtable}
}

\parbox{.5\linewidth}{
\centering
\begin{subtable}[c]{.3\textwidth}
\centering
\resizebox{0.8\textwidth}{10mm}{
\begin{tabular}{|l|l|l|l|}
\hline
\textbf{r(0)}                       & \textbf{p1} & \textbf{p2} & \textbf{p3} \\ \hline
\textbf{10\textasciicircum{}(-5)}   & 0.883       & 0.058       & 0.058       \\ \hline
\textbf{10\textasciicircum{}(-4.8)} & 0.862       & 0.069       & 0.069       \\ \hline
\textbf{10\textasciicircum{}(-4.6)} & 0.845       & 0.078       & 0.078       \\ \hline
\textbf{10\textasciicircum{}(-4.4)} & 0.818       & 0.091       & 0.091       \\ \hline
\textbf{10\textasciicircum{}(-4.2)} & 0.795       & 0.102       & 0.102       \\ \hline
\textbf{10\textasciicircum{}(-4)}   & 0.767       & 0.116       & 0.116       \\ \hline
\end{tabular}}
\caption{$\rezprob(0)=10^{-5} \sim 10^{-4}$}
\end{subtable}

\begin{subtable}[c]{.3\textwidth}
\centering
\resizebox{0.8\textwidth}{10mm}{
\begin{tabular}{|l|l|l|l|}
\hline
\textbf{r(0)}                       & \textbf{p1} & \textbf{p2} & \textbf{p3} \\ \hline
\textbf{10\textasciicircum{}(-3)}   & 0.621       & 0.225       & 0.155       \\ \hline
\textbf{10\textasciicircum{}(-2.8)} & 0.596       & 0.243       & 0.160       \\ \hline
\textbf{10\textasciicircum{}(-2.6)} & 0.578       & 0.259       & 0.163       \\ \hline
\textbf{10\textasciicircum{}(-2.4)} & 0.566       & 0.271       & 0.163       \\ \hline
\textbf{10\textasciicircum{}(-2.2)} & 0.558       & 0.282       & 0.159       \\ \hline
\textbf{10\textasciicircum{}(-2)}   & 0.559       & 0.290       & 0.151       \\ \hline
\end{tabular}}
\caption{$\rezprob(0)=10^{-3} \sim 10^{-2}$}
\end{subtable}

\begin{subtable}[c]{.3\textwidth}
\centering
\resizebox{0.8\textwidth}{10mm}{
\begin{tabular}{|l|l|l|l|}
\hline
\textbf{r(0)}                        & \textbf{p1} & \textbf{p2} & \textbf{p3} \\ \hline
\textbf{10\textasciicircum{}(-0.1)}  & 1.00        & 0.00        & 0.00        \\ \hline
\textbf{10\textasciicircum{}(-0.08)} & 1.00        & 0.00        & 0.00        \\ \hline
\textbf{10\textasciicircum{}(-0.06)} & 1.00        & 0.00        & 0.00        \\ \hline
\textbf{10\textasciicircum{}(-0.04)} & 1.00        & 0.00        & 0.00        \\ \hline
\textbf{10\textasciicircum{}(-0.02)} & 1.00        & 0.00        & 0.00        \\ \hline
\textbf{10\textasciicircum{}(0)}     & 1.00        & 0.00        & 0.00        \\ \hline
\end{tabular}}
\caption{$\rezprob(0)=10^{-0.1} \sim 1$}
\end{subtable}
}
}
\caption{The channel selection probability vector $\boldsymbol{p}$  that minimizes the ETTR for $N=3$ and various $\rezprob(0)$'s  (chosen in log-scale from $10^{-5}$ to $0$) with $\rho = 0.1$ in (a),(b),(c) and $\rho = 0.9$ in (d),(e),(f). }
\label{table:p_dis}
\end{table}

The asymptotic $(1+\epsilon)$-approximation solution in \rthe{twomain} and the numerical results in Table \ref{table:p_dis} motivate us to propose a local search algorithm, called the
{\em improved uniform selection policy} in Algorithm \ref{alg:multipleb}.
The basic idea is to transform a probability vector $\boldsymbol{u}$ (starting from the uniform selection probability vector) to another one ${\boldsymbol{u}}^\prime$ so that ${\boldsymbol{u}} \prec {\boldsymbol{u}}^\prime$.
Since $g({\boldsymbol{u}})$ is Schur concave and $f({\boldsymbol{u}})$ is Schur convex,
the majorization ordering between $\boldsymbol{u}$ and ${\boldsymbol{u}}^\prime$ then ensures that
 $g({\boldsymbol{u}}) \ge  g({\boldsymbol{u}}^\prime)$ and $f({\boldsymbol{u}}) \le  f({\boldsymbol{u}}^\prime)$.
Then we carry out successive transformations until there is no further improvement of ETTR.
In the last step of Algorithm \ref{alg:multipleb}, we compare the ETTR found by local search with the ETTR of the single selection policy and then choose the better one. By doing so, Algorithm \ref{alg:multipleb} is also a $\min[M,N]$-approximation solution.

\begin{algorithm}
\caption{The improved uniform selection policy}\label{alg:multipleb}
\noindent {\bf Input}: {The joint probability mass function $q(\boldsymbol{x})$ for the set of $\{X_1(0),X_2(0))\ldots,X_N(0)\}$ exchangeable random variables.
}

\noindent {\bf Output}: {The channel selection probabilities $p_i$, $i=1,2, \ldots, N$.}

\noindent 1: Let $\boldsymbol{p}$ be channel selection probability vector of the uniform selection policy, i.e., $\boldsymbol{p}=(p_1,p_2,\ldots,p_N)$ with $p_i=1/N$, $i=1,2,\ldots,N$.

\noindent 2: Let $\boldsymbol{u}=(u_1,u_2,\ldots,u_N)$ and
 compute $u_i=\frac{p_i^2}{\sum_{j=1}^N p_j^2}$, $i=1,2,\ldots, N$.

\noindent 3: Let  $\ell=1$. Choose a step size $\Delta$.

\noindent 4: Update $\boldsymbol{u}$ by assigning $u_\ell \mapsto u_\ell+\Delta*(N-\ell)$ and $u_i \mapsto u_i-\Delta$, $i=\ell+1,\ldots, N$.

\noindent 5: Compute the ETTR by using $g({\boldsymbol{u}})f({\boldsymbol{u}})$ with $g({\boldsymbol{u}})$ in \req{change4444} and $f({\boldsymbol{u}})$ in \req{ettr3377}.

\noindent 6: If there is an improvement of the ETTR, repeat from Step 4. Otherwise update $\ell \mapsto \ell+1$.

\noindent 7: While $\ell<N$, repeat from Step 4.

\noindent 8: Otherwise  compute $p_i =\frac{\sqrt{u_i}}{\sum_{j=1}^N \sqrt{u_j}}$.

\noindent 9: If the ETTR is smaller than that of the single selection policy, output $p_i,i=1,2,\ldots,N$. Otherwise output $\boldsymbol{p}$ of the single selection policy.

\end{algorithm}

The grid search is a simplified version of the ``exhaustive'' search, and
it should be close to the optimal solution if the resolution size $\Delta$ is set to be very small.
Clearly, the number of channel selection probability vectors that need to be searched by the grid search is $O((\frac{1}{\Delta})^{N-1})$.
As such, if the resolution size $\Delta$ is very small,
then its computational cost is very high. For our experiments, we are only able to conduct the grid search with $\Delta=0.001$ for $N=4$ due to its
large computational cost. On the other hand, the step size $\Delta$ of Algorithm \ref{alg:multipleb} is also set to be $0.001$.
Note that the number of channel selection probability vectors that need to be searched by Algorithm \ref{alg:multipleb}  is only $O(\frac{N}{\Delta})$.
In Table \ref{table:ETTR_algcom_2},  we show the comparison results for ETTR  between Algorithm \ref{alg:multipleb} (A1) and the grid search (Grid) when there are four channels ($N=4$) and $\rezprob(0)=0.1,0.01,0.001$ and $0.0001$, respectively.
 Note that the ETTR results obtained by  Algorithm \ref{alg:multipleb}  are
very close to those from the grid search (and it is better when $\rho=0.4$ and $\rezprob(0)=0.001$). These numerical results suggest that Algorithm \ref{alg:multipleb} might be a
more scalable and effective alternative than the grid search.

\begin{table}
\makebox[0.87\linewidth][c]{
\parbox{.5\linewidth}{
\centering
\begin{subtable}[c]{.3\textwidth}
\centering
\resizebox{0.6\textwidth}{14mm}{
\begin{tabular}{|l|l|l|}
\hline
$\boldsymbol{\rho}$ & \textbf{A1} &  \textbf{Grid} \\ \hline
\textbf{0.1} & 9.10        & 9.10          \\ \hline
\textbf{0.2} & 8.20        & 8.20          \\ \hline
\textbf{0.3} & 7.30        & 7.30          \\ \hline
\textbf{0.4} & 6.40        & 6.40          \\ \hline
\textbf{0.5} & 5.50        & 5.50          \\ \hline
\textbf{0.6} & 4.60        & 4.60          \\ \hline
\textbf{0.7} & 3.70        & 3.70          \\ \hline
\textbf{0.8} & 2.80        & 2.80          \\ \hline
\textbf{0.9} & 1.90        & 1.90          \\ \hline
\end{tabular}}
\caption{$\rezprob(0)=0.1$}
\end{subtable}

\begin{subtable}[c]{.3\textwidth}
\centering
\resizebox{0.6\textwidth}{14mm}{
\begin{tabular}{|l|l|l|}
\hline
$\boldsymbol{\rho}$ & \textbf{A1} & \textbf{Grid} \\ \hline
\textbf{0.1} & 90.10       & 90.10         \\ \hline
\textbf{0.2} & 80.20       & 80.20         \\ \hline
\textbf{0.3} & 70.30       & 70.30         \\ \hline
\textbf{0.4} & 49.31       & 49.31         \\ \hline
\textbf{0.5} & 30.35       & 30.35         \\ \hline
\textbf{0.6} & 17.24       & 17.24         \\ \hline
\textbf{0.7} & 9.60        & 9.60          \\ \hline
\textbf{0.8} & 5.93        & 5.90          \\ \hline
\textbf{0.9} & 3.81        & 3.61          \\ \hline
\end{tabular}}
\caption{$\rezprob(0)=0.01$}
\end{subtable}
}

\parbox{.5\linewidth}{
\begin{subtable}[c]{.3\textwidth}
\centering
\resizebox{0.6\textwidth}{14mm}{
\begin{tabular}{|l|l|l|}
\hline
$\boldsymbol{\rho}$ & \textbf{A1} & \textbf{Grid} \\ \hline
\textbf{0.1} & 900.10      & 900.10        \\ \hline
\textbf{0.2} & 670.26      & 670.27        \\ \hline
\textbf{0.3} & 442.57      & 442.37        \\ \hline
\textbf{0.4} & 268.86      & 268.86        \\ \hline
\textbf{0.5} & 148.44      & 148.44        \\ \hline
\textbf{0.6} & 72.44       & 72.44         \\ \hline
\textbf{0.7} & 30.03       & 30.03         \\ \hline
\textbf{0.8} & 10.58       & 10.57         \\ \hline
\textbf{0.9} & 4.34        & 4.27          \\ \hline
\end{tabular}}
\caption{$\rezprob(0)=0.001$}
\end{subtable}

\begin{subtable}[c]{.3\textwidth}
\centering
\resizebox{0.6\textwidth}{14mm}{
\begin{tabular}{|l|l|l|}
\hline
$\boldsymbol{\rho}$ & \textbf{A1} & \textbf{Grid} \\ \hline
\textbf{0.1} & 8095.36     & 7886.00       \\ \hline
\textbf{0.2} & 5276.89     & 5267.19       \\ \hline
\textbf{0.3} & 3288.77     & 3274.10       \\ \hline
\textbf{0.4} & 1878.71     & 1878.76       \\ \hline
\textbf{0.5} & 973.35      & 973.34        \\ \hline
\textbf{0.6} & 438.26      & 437.99        \\ \hline
\textbf{0.7} & 159.70      & 159.70        \\ \hline
\textbf{0.8} & 41.36       & 41.36         \\ \hline
\textbf{0.9} & 6.70        & 6.64          \\ \hline
\end{tabular}}
\caption{$\rezprob(0)=0.0001$}
\end{subtable}
}
}
\caption{Comparisons of  ETTR for  Algorithm \ref{alg:multipleb} (A1)  and the grid search (Grid) for $4$ channels with $\rezprob(0)=0.1,0.01,0.001$ and $0.0001$.}
\label{table:ETTR_algcom_2}
\end{table}

\bsec{A general time-varying channel model}{general}

In this section, we consider a general time varying channel model.

\bdefin{general}{\bf (General Time-Varying Channel Model)}
For the general time-varying channel model, the sequence of random vectors
$\{\boldsymbol{X}(t)=(X_1(t),X_2(t),\ldots, X_N(t)), t\ge 0\}$ are assumed to be {\em stationary}, i.e.,
its joint distribution is invariant with respect to any time shift. Moreover, the joint probability mass function at time 0 is assumed to be
$$\pr(\boldsymbol{X}(0)=\boldsymbol{x})=q(\boldsymbol{x}),$$
where $\boldsymbol{x}=(x_1, x_2, \ldots, x_N)$.
\edefin

Clearly, both the fast time-varying channel  channel in \rdef{fast} and the slow time-varying channel channel model in \rdef{slow} are special cases of this general channel model (as a sequence of i.i.d. random vectors is stationary and a sequence of identical random vector is also stationary).


Let ${T(\boldsymbol{p})}$ (resp. ${T_f(\boldsymbol{p})}$, ${T_s(\boldsymbol{p})}$) be the ETTR  for the blind rendezvous policy with the channel selection probabilities
$\boldsymbol{p}=(p_1,p_2, \ldots, p_N)$ in the general (resp. fast and slow) time-varying channel model.

\bthe{general}
The ETTR in  the general time-varying channel model is upper bounded by the ETTR of the   slow time-varying channel model, i.e.,
\beq{gen1111}
\ex[T(\boldsymbol{p})] \le \ex[T_s(\boldsymbol{p})].
\eeq
\ethe

Our proof for \rthe{general} is based on the stochastic ordering (see e.g., Chapter 9 of \cite{Ross1996}).
We say that a random variable $X$ is {\em stochastically larger} than another random variable $Y$, denoted by $X \ge_{st} Y$, if
$$\pr (X >t) \ge \pr (Y >t)$$
for all $t$. It is well-known that $X \ge_{st} Y$ if and only if $ \ex[f(X)] \ge \ex[f(Y)]$ for all increasing function $f$ (Proposition 9.1.2 of \cite{Ross1996}). As a result, $\ex[X] \ge \ex[Y]$ if $X \ge_{st} Y$.

For the proof of \rthe{general}, we need the following lemma. 

\blem{Holder}
Suppose that a set of $t$ nonnegative random variables  $Z(s), s=1, 2, \ldots, t$ have a common distribution as that of the random variable $Z(0)$.
Then
\beq{Holder1111}
\ex[\prod_{s=1}^t Z(s)] \le \ex[Z(0)^t].
\eeq
\elem

\bproof
We prove this by the H\"older inequality, i.e., for any two random variables
$1/p+1/q=1$, and $p,q \ge 1$,
$$\ex[|X Y|] \le (\ex[|X|^p])^{1/p} (\ex[|X|^q])^{1/q}.$$
For $t=2$, it follows directly from the H\"older inequality and the nonnegativity of the $t$ random variables that
$$\ex[Z(1) Z(2)] \le (\ex[Z(1)^2])^{1/2} (\ex[Z(2)^2])^{1/2}.$$
Since these random variables are identically distributed as that of the random variable $Z(0)$, we then have
$$\ex[Z(1) Z(2)] \le \ex[Z(0)^2].$$
Now assume that the inequality in \req{Holder1111} holds for $t-1$ as the induction hypothesis.
Choose $q=t$ and $p=t/(t-1)$.
It then follows from the H\"older inequality and the nonnegativity of the $t$ random variables that
\bearn
&&\ex[\prod_{s=1}^t Z(s)] \\
&&=\ex[ (\prod_{s=1}^{t-1} Z(s))(Z(t))] \\
&&\le \Big (\ex[(\prod_{s=1}^{t-1} Z(s))^{t/(t-1)}] \Big )^{(t-1)/t} \Big (\ex[(Z(t)^t] \Big)^{1/t} \\
&&= \Big (\ex[\prod_{s=1}^{t-1} Z(s)^{t/(t-1)}]\Big )^{(t-1)/t} \Big (\ex[(Z(0)^t]\Big)^{1/t} .
\eearn
Using the induction hypothesis yields
\bearn
&&\ex[\prod_{s=1}^t Z(s)] \\
&&\le \Big (\ex[Z(0)^{t}]\Big)^{(t-1)/t} \Big(\ex[(Z(0)]^t]\Big)^{1/t} =\ex[Z(0)^t].
\eearn
\eproof

\bproof(\rthe{general})
Note that given the channel states $X(s)=x(s)=(x_1(s), x_2(s), \ldots, x_N(s))$ at time $s$, whether the two users will rendezvous at time $s$ is an independent Bernoulli random variable with parameter $\sum_{i=1}^{N}p_i^2\rezprob(x_i(s))$. Thus,
\bear{gen2222}
&&\pr(T(\boldsymbol{p}) > t | X(s)=x(s), 1 \le s \le t) \nonumber\\
&& =\prod_{s=1}^t (1-\sum_{i=1}^{N}p_i^2\rezprob(x_i(s))).
\eear
Unconditioning on the event $\{X(s)=x(s), 1 \le s \le t\}$ yields
\bear{gen3333}
&&\pr(T(\boldsymbol{p}) > t)\nonumber\\
&&=\ex[\prod_{s=1}^t (1-\sum_{i=1}^{N}p_i^2\rezprob(X_i(s)))] \nonumber \\
&&=\ex[\prod_{s=1}^t Z(s)],
\eear
where
\beq{gen4444}
Z(s)=1-\sum_{i=1}^{N}p_i^2 \rezprob(X_i(s)).
\eeq
On the other hand, for the slow time-varying model,
we have
\bear{gen3355}
&&\pr(T_s(\boldsymbol{p}) > t | X(s)=x(s), 1 \le s \le t) \nonumber\\
&&=\pr(T_s(\boldsymbol{p}) > t | X(0)=x(0)) \nonumber\\
&&=(1-\sum_{i=1}^{N}p_i^2\rezprob(x_i(0)))^t
\eear
and thus
\beq{gen3366}
\pr(T_s(\boldsymbol{p}) > t )= \ex[(1-\sum_{i=1}^{N}p_i^2\rezprob(X_i(0)))^t]=\ex[Z(0)^t].
\eeq
It then follows from \rlem{Holder} that for all $t$
\bear{gen5555}
\pr(T(\boldsymbol{p}) > t)&=&\ex[\prod_{s=1}^t Z(s)]\nonumber\\
&\le& \ex[ Z(0)^t]=\pr(T_s(\boldsymbol{p}) > t).
\eear
This implies that $T_s(\boldsymbol{p})$ is stochastically larger than $T(\boldsymbol{p})$ and thus
$$\ex[T(\boldsymbol{p})] \le \ex[T_s(\boldsymbol{p})].$$
\eproof

\bsec{A Markov channel model with two states}{gtwo}

In this section, we consider a Markov channel model with two states.

As in \rsec{two}, we assume that the states of these $N$ channels are {\em independent and identically distributed}. The probability that the $i^{th}$ channel is in the good (resp. bad) state is $\rho_i$ (resp. $1-\rho_i$) for some $0 \le \rho_i \le 1$.
As such,  we have the following stationary joint distribution for the channel states
\bear{twojoint111g}
&&\pr (X_1(t)=x_1, X_2(t)=x_2, \ldots, X_N(t)=x_N) \nonumber\\
&&
=q(x_1,x_2, \ldots, x_N)=\prod_{i=1}^N \rho_i^{x_i} (1-\rho_i)^{1-x_i},
\eear
where $\rho_i$ (resp. $(1-\rho_i)$ is the probability of being in state 1 (resp. 0), and $x_i$ (with the value being 0 or 1) is the state of channel $i$.
For the $i^{th}$ channel, its channel state is characterized by a Markov chain with the transition probabilities:
\bear{tran1111}
&&\pr (X_i(t+1)=1 | X_i(t)=1)=\pon, \label{eq:tran1111aa}\\
&&\pr (X_i(t+1)=0 | X_i(t)=1)=1-\pon, \label{eq:tran1111ab}\\
&&\pr (X_i(t+1)=0 | X_i(t)=0)=\poff, \label{eq:tran1111bb} \\
&&\pr (X_i(t+1)=1 | X_i(t)=0)=1-\poff , \label{eq:tran1111ba}
\eear
where $0 < \pon, \poff <1$.
Clearly, we have
$$\rho_i=\pr( X_i(t)=1)=\frac{1- \poff }{(1-\pon) +(1-\poff)}.$$
Note that
$$\mbox{Var}[X_i(t+1)]=\mbox{Var}[X_i(t)]=\rho_i (1- \rho_i)$$
and thus
the correlation coefficient between $X_i(t+1)$ and $X_i(t)$, denoted by $\omega^{(i)}$, is
\bear{tran1155}
&&\frac{\ex [X_i(t+1) X_i(t)] -\ex[X_i(t+1)] \ex [X_i(t)]}{\sqrt{\mbox{Var}[X_i(t+1)] \mbox{Var}[X_i(t)] }}\nonumber\\
&&=\frac{\rho_i \pon -\rho_i^2}{\rho_i (1- \rho_i)} \nonumber \\
&&=\pon+\poff -1.
\eear
We say that the Markov chain  $\{X_i(t), t \ge 0\}$ is positively correlated  if $\omega^{(i)} \ge 0$.

\bsubsec{An ETTR lower bound for positively correlated Markov chains}{lowerM}

In this section,
we show that the ETTR of the Markov chain model is lowered bounded by that of the   fast time-varying channel model
if the two-state Markov chains are positively correlated.

\bthe{general2}
Consider the $N$ independent Markov channel model with two states in this section.
Let $\omega^{(i)}=\pon+\poff -1$.
If $\omega^{(i)} \ge 0$ for all $i$, then
the ETTR in  the Markov channel model is lower bounded by the ETTR of the   fast time-varying channel model, i.e.,
\beq{gen11112}
\ex[T(\boldsymbol{p})] \ge \ex[T_f(\boldsymbol{p})].
\eeq
\ethe

Our proof for \rthe{general2} is based on the Lorentz inequality and the coupling of two-state Markov chains that was previously used in \cite{Chang90}.
 A function $f:( x_1, \ldots , x_n ) \to {\cal R}$ is supermodular (or L-superadditive) if for any
  $1 \le
 i < j \le n$ and any nonnegative $y_1$, $y_2$ the following inequality holds:
\bearn
&&f(\ldots , x_i +y_1 , \ldots , x_j +y_2 , \ldots )\\&&\quad +f(\ldots , x_i,
 \ldots , x_j, \ldots ) \\
 &&\ge f(\ldots , x_i +y_1 , \ldots , x_j, \ldots ) \\&&\quad+f(\ldots , x_i,
 \ldots , x_j+y_2 , \ldots ).
 \eearn
 According to Marshall and
 Olkin \cite{MarOl} (Chapter 6), the condition above is equivalent to $\frac{\partial^2 f}{\partial x_i \partial x_j} \ge 0$ for all $i$, $j$ ($i \ne j$) if the function is twice differentiable.

An interesting property of supermodular functions is the Lorentz inequality.
  Let $Z_j
 ,j=1,\ldots ,n$, be a sequence of not necessarily independent r.v.'s with a
 common distribution. Then for any supermodular function $f$ (see \cite{Tchen80} Theorem 5A and \cite{Rolski86}
 Lemma 5),
 $$Ef(Z_1 , \ldots   , Z_n ) \le Ef(Z_1 ,\ldots , Z_1 ).$$

 An immediate extension of the Lorentz inequality to independent random vectors
  is
 presented in the following lemma.

\blem{Lorentz}
 (The Lorentz inequality for independent random vectors \cite{Chang90}) Let $Z^{(i)}=(Z^{(i)}_1, \ldots,
 Z^{(i)}_{n_i} )$, $i =1 , \ldots ,m$,  be $m$ independent random vectors. If the
  r.v.'s
 $Z^{(i)}_j$ , $j=1 \ldots n_i$, have a common distribution
 $F^{(i)} (x) =\pr( Z^{(i)}_j < x)$, then for any supermodular function $f$,
$$\ex[f(Z^{(1)}_1, \ldots,Z^{(1)}_{n_1}, Z^{(2)}_1, \ldots,Z^{(2)}_{n_2},
  \ldots
 ,Z^{(m)}_1, \ldots,Z^{(m)}_{n_m})]$$
$$\le \ex [f(Z^{(1)}_1, \ldots,Z^{(1)}_{1},
 Z^{(2)}_1, \ldots,Z^{(2)}_{1}, \ldots
 ,Z^{(m)}_1, \ldots,Z^{(m)}_{1})].
$$
 \elem

 \bproof(\rthe{general2})
 Since $\omega^{(i)} \ge 0$, one can construct such a stationary Markov chain
by generating two independent
  sequences of i.i.d.
 Bernoulli r.v.'s $\{ \alpha^{(i)}(t) , \beta^{(i)}(t), t \ge 0 \}$ with parameters
 $\omega^{(i)}$ and $\rho$.
  Construct $X_i(t)$ according to the following rules:
  \begin{description}
 \item[(i)] $X_i(0) = \beta^{(i)}(0) $.
 \item[(ii)] $X_i(t+1) = X_i(t)$ if $\alpha^{(i)}(t) =1$.
 \item[(iii)] $X_i(t+1)  =\beta^{(i)}(t+1)$ if $\alpha^{(i)}(t)=0$.
\end{description}
It is easy to verify that such a construction is a stationary Markov chain with the transition probabilities specified in
\req{tran1111aa}-\req{tran1111ba}.

As discussed in the proof of \rthe{general}, given the channel states $X(s)=x(s)=(x_1(s), x_2(s), \ldots, x_N(s))$ at time $s$, whether the two users will rendezvous at time $s$ is an independent Bernoulli random variable with parameter $\sum_{i=1}^{N}p_i^2\rezprob(x_i(s))$. Thus,
\bear{gen2222l}
&&\pr(T(\boldsymbol{p}) > t | X(s)=x(s), 1 \le s \le t) \nonumber\\
&& =\prod_{s=1}^t (1-\sum_{i=1}^{N}p_i^2\rezprob(x_i(s))).
\eear
and
\bear{gen3333l}
&&\pr(T(\boldsymbol{p}) > t)\nonumber\\
&&=\ex[\prod_{s=1}^t (1-\sum_{i=1}^{N}p_i^2\rezprob(X_i(s)))].
\eear
It is easy to see that $\pr(T(\boldsymbol{p}) > t)$ is a supermodular function of $X_i(s)$, $1 \le s \le t$, $1 \le i \le N$.
Also, as the fast time-varying channel model corresponds to the case that $\omega^{(i)}=0$. This then implies that
$\alpha^{(i)}(t)=0$ and $X_i(t+1)=\beta^{(i)}(t+1)$ for all $t$ in the fast time-varying channel model.
As a direct result of the Lorentz inequality in \rlem{Lorentz} and the construction of the two-state Markov chains,
we then have
$$\pr(T(\boldsymbol{p}) > t) \ge \pr(T_f(\boldsymbol{p}) > t)=(1-\sum_{i=1}^{N}p_i^2\ex[\rezprob(X_i(0))])^t.$$
This implies that $T_f(\boldsymbol{p})$ is stochastically smaller than $T(\boldsymbol{p})$ and thus
$$\ex[T(\boldsymbol{p})] \ge \ex[T_f(\boldsymbol{p})].$$
\eproof

\bsubsec{Approximation solutions}{approxM}

In this section, we extend the two approximation solutions in \rthe{approx} for the slow time-varying channel model to the two-state Markov chain channel model.
We show that the ETTR of the uniform selection policy is very insensitive to the underlining two-state Markov channel model when the number of channels $N$ is very large. As such, it is an asymptotic $N$-approximation solution for the ETTR minimization problem in \req{opti0011}. On the other hand, the single selection policy is an $M$-approximation solution with the same constant $M$ defined in
\req{approx0022}.

\bthe{insensitivity}
Consider the $N$ independent Markov channel model with two states in this section.
Suppose that  (i) $\omega^{(i)} \ge 0$ for all $i$,  (ii) $\rho_i =\rho$ for all $i$, and (iii) $r(0)>0$.
\begin{description}
\item[(i)] The ETTR of the uniform selection policy  that uses the channel selection probability vector $\boldsymbol{p^u}=(1/N,1/N, \ldots, 1/N)$ has the following asymptotic result:
\beq{insen1111}
\lim_{N \to \infty }\frac{\ex[T(\boldsymbol{p^u})]}{N} =\frac{1}{\rho r(1)+(1-\rho)r(0)} .
\eeq
Moreover, the uniform selection policy is an asymptotic $N$-approximation solution for the ETTR minimization problem in \req{opti0011} when $N \to \infty$,
i.e., for any channel selection probability vector $\boldsymbol{p}$,
\beq{approxM1111}
\lim_{N \to \infty}\frac{\ex[T(\boldsymbol{p^u})]}{N \ex[T(\boldsymbol{p})]} \le 1 .
\eeq
\item [(ii)] The single selection policy that uses the channel selection probability vector $\boldsymbol{p^s}=(1,0, \ldots, 0)$ is an $M$-approximation solution for the ETTR minimization problem in \req{opti0011},
 i.e., for any channel selection probability vector $\boldsymbol{p}$,
\beq{approxM2222}
\frac{\ex[T(\boldsymbol{p^s})]}{\ex[T(\boldsymbol{p})]} \le M ,
\eeq
where the constant
\beq{approxM0022}
M=\frac{\rho\frac{1}{r(1)}+(1-\rho){\frac{1}{r(0)}}}{\frac{1}{\rho r(1)+(1-\rho)r(0)}}.
\eeq
\end{description}
\ethe

\bproof
\noindent (i)
From the ETTR upper bound in \req{gen1111} and the ETTR lower bound \req{gen11112}, we have that
\beq{insen2222}
\ex[T_f(\boldsymbol{p^u})] \le \ex[T(\boldsymbol{p^u})] \le \ex[T_s(\boldsymbol{p^u})].
\eeq
Since we assume that $\rho_i=\rho$ for all $i$, it then follow from the ETTR for the fast time-varying channel model in \req{fastETTR} that
\bear{insen3333}
\ex[T_f(\boldsymbol{p^u})]  &=&1 / \sum_{i=1}^{N}(1/N)^2 \ex[\rezprob(X_i(0))]\nonumber \\
&=&
\frac{N}{\rho r(1)+(1-\rho)r(0)}.
\eear

On the other hand, we have from the ETTR for the slow time-varying channel model in \req{ettr2222} that
\bear{insen4444}
\ex[T_s(\boldsymbol{p^u})] & = &\ex \Big [\frac{1}{\sum_{i=1}^{N}\rezprob(X_i(0))(1/N)^2}\Big ]\nonumber \\
&=&N \ex \Big [\frac{N}{\sum_{i=1}^{N}\rezprob(X_i(0))}\Big ].
\eear
As the $N$ Markov chains are independent and $\rho_i=\rho$ for all $i$,
the $N$ random variables $\rezprob(X_i(0))$, $i=1,2, \ldots, N$, are independent and identically distributed.
We then have from the strong law of large numbers that
\beq{insen4455}\lim_{N \to \infty}\frac{1}{N}{\sum_{i=1}^{N}\rezprob(X_i(0))}=\rho r(1)+(1-\rho)r(0),\quad a.s.
\eeq
Since we assume that $r(0)>0$, the sequence of random variables
$\{\frac{N}{\sum_{i=1}^{N}\rezprob(X_i(0))}, N \ge 1\}$, are all bounded between $1$ and $1/r(0)$.
It then follows from the bounded convergence theorem (for the exchange of the limit and the expectation) and the strong law in \req{insen4455} that
\bear{insen5555}
&&\lim_{N \to \infty}\frac{\ex[T_s(\boldsymbol{p^u})]}{N} \nonumber\\
&&=\lim_{N \to \infty}\ex \Big [\frac{N}{\sum_{i=1}^{N}\rezprob(X_i(0))}\Big ]\nonumber \\
&&=\ex \Big [\lim_{N \to \infty}\frac{N}{\sum_{i=1}^{N}\rezprob(X_i(0))}\Big ]\nonumber \\
&&=\frac{1}{\rho r(1)+(1-\rho)r(0)}.
\eear
The result in \req{insen1111} then follows from \req{insen2222}, \req{insen3333} and \req{insen5555}.

 From the ETTR lower bound \req{gen11112} and the result that the single channel selection  policy is optimal for the fast time-varying channel model in \rthe{fast}, we have
\bear{insen6666}
&&\ex[T(\boldsymbol{p})] \ge \ex[T_f(\boldsymbol{p})] \nonumber \\
&&\ge \frac{1}{\ex[r(X_1(0))]}=\frac{1}{\rho r(1)+(1-\rho)r(0)}.
\eear
Using this and \req{insen1111} yields \req{approxM1111}.

\noindent (ii) As shown in \req{insen6666}, we have
\beq{insen6666S}
\ex[T(\boldsymbol{p})] \ge  \frac{1}{\rho r(1)+(1-\rho)r(0)}.
\eeq
On the other hand, we have from the ETTR upper bound in \req{gen1111}  that for the single channel selection policy
\beq{insen2222S}
 \ex[T(\boldsymbol{p^s})] \le \ex[T_s(\boldsymbol{p^s})]={\rho\frac{1}{r(1)}+(1-\rho){\frac{1}{r(0)}}}.
\eeq
The result in \req{approxM2222} then follows from the two inequalities in \req{insen6666S} and \req{insen2222S}.
\eproof

\bsec{Conclusion}{conclusion}

In this paper, we considered the multichannel rendezvous problem in CRNs where the probability that
two users hopping on the same channel have a successful rendezvous  is a function of channel states.
 We first considered two channel models: (i) the fast time-varying channel model, and (ii) the slow time-varying channel model.
Among the classes of the  blind rendezvous policies that randomly hop on channels according to certain channel selection probabilities, we showed
the optimal channel selection policy that minimizes the ETTR is the single selection policy that  hops on the ``best'' channel all the time in the fast time-varying channel model. However, this is not the case for the slow time-varying channel model.
The intuition behind this is that the ``best'' channel  might be in fact in a very bad state with a very low rendezvous probability. This then leads to a very large ETTR and it is preferable to having  nonzero probabilities to hop on the other channels.
For the slow time-varying channel model, we used the majorization ordering to derive various bounds and approximation algorithms
when the channel states are exchangeable/i.i.d. random variables. By conducting extensive numerical experiments,
we also verified the effectiveness of our approximation algorithms.

We then extended our results to general channel models, where  the joint distribution of the channel states is only assumed to be stationary in time.
We showed that its ETTR is upper bounded by the ETTR of the slow time-varying channel model.
On the other hand, we also generalized the i.i.d. two-state model to a two-state Markov chain model and
showed that if the two-state Markov chain is positively correlated, then
its ETTR  is lower bounded by the ETTR of the  fast time-varying channel model.
Based on both the lower bound and the upper bound, we further showed that
the uniform selection policy is an asymptotic $N$-approximation solution and the
single selection policy is an $M$-approximation solution for any positively correlated two-state Markov chains.

There are several possible extensions of this work: (i) heterogeneous environments: here we assume that the joint distribution of channel states is the same for the two users. Such a homogeneous assumption is valid if the two users are close to each other in a CRN. On the other hand, if the two users are far apart, then they might have different joint distributions of channel states. (ii) partial observable channel states: here we assume that the channel states are not observable by the two users and thus the two users cannot ``learn'' from failed rendezvous. For the slow time-varying channel model, it might be possible for users to ``learn'' the state of a channel by using reinforcement learning \cite{learning}.
However, it is probably not worth the trouble if the ETTR is much shorter than the learning time.

\begin{IEEEbiography}[{\includegraphics[width=1in,height=1.25in,clip,keepaspectratio]{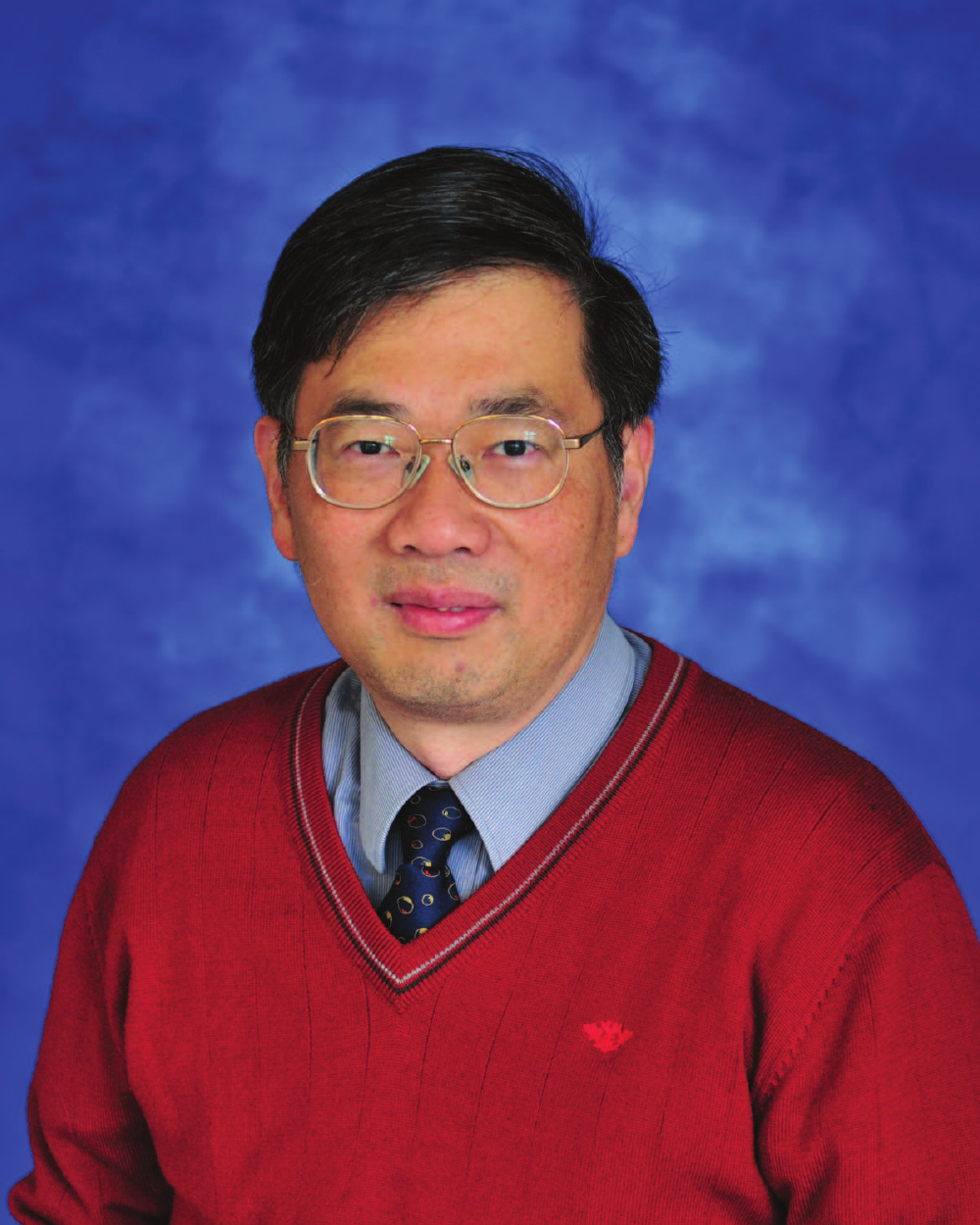}}]
{Cheng-Shang Chang}
(S'85-M'86-M'89-SM'93-F'04)
received the B.S. degree from National Taiwan
University, Taipei, Taiwan, in 1983, and the M.S.
and Ph.D. degrees from Columbia University, New
York, NY, USA, in 1986 and 1989, respectively, all
in electrical engineering.

From 1989 to 1993, he was employed as a
Research Staff Member with the IBM Thomas J.
Watson Research Center, Yorktown Heights, NY,
USA. Since 1993, he has been with the Department
of Electrical Engineering, National Tsing Hua
University, Taiwan, where he is a Tsing Hua Distinguished Chair Professor. He is the author
of the book Performance Guarantees in Communication Networks (Springer,
2000) and the coauthor of the book Principles, Architectures and Mathematical
Theory of High Performance Packet Switches (Ministry of Education, R.O.C.,
2006). His current research interests are concerned with network science, big data analytics,
mathematical modeling of the Internet, and high-speed switching.

Dr. Chang served as an Editor for Operations Research from 1992 to 1999,
an Editor for the {\em IEEE/ACM TRANSACTIONS ON NETWORKING} from 2007
to 2009, and an Editor for the {\em IEEE TRANSACTIONS
ON NETWORK SCIENCE AND ENGINEERING} from 2014 to 2017. He is currently serving as an Editor-at-Large for the {\em IEEE/ACM
TRANSACTIONS ON NETWORKING}. He is a member of IFIP Working
Group 7.3. He received an IBM Outstanding Innovation Award in 1992, an
IBM Faculty Partnership Award in 2001, and Outstanding Research Awards
from the National Science Council, Taiwan, in 1998, 2000, and 2002, respectively.
He also received Outstanding Teaching Awards from both the College
of EECS and the university itself in 2003. He was appointed as the first Y. Z.
Hsu Scientific Chair Professor in 2002. He received the Merit NSC Research Fellow Award from the
National Science Council, R.O.C. in 2011. He also received the Academic Award in 2011 and the National Chair Professorship in 2017 from
the Ministry of Education, R.O.C. He is the recipient of the 2017 IEEE INFOCOM Achievement Award.
\end{IEEEbiography}

\begin{IEEEbiography}
[{\includegraphics[width=1in,height=1.25in,clip,keepaspectratio]{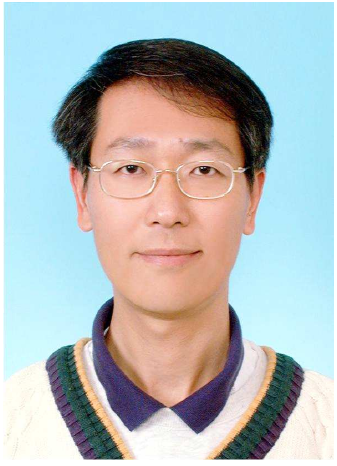}}]
{Duan-Shin Lee}(S'89-M'90-SM'98) received the B.S. degree from National Tsing Hua
University, Taiwan, in 1983, and the MS and Ph.D. degrees from
Columbia University, New York, in 1987 and 1990, all in electrical
engineering.  He worked as a research staff member at the C\&C Research Laboratory
of NEC USA, Inc. in Princeton, New Jersey from 1990 to 1998.  He joined the
Department of Computer Science of National Tsing Hua University in Hsinchu,
Taiwan, in 1998.  Since August 2003, he has been a professor.  He received
a best paper award from the Y.Z. Hsu Foundation in 2006.  He served as
an editor for the Journal of Information Science and Engineering between
2013 and 2015.  He is currently an editor for Performance Evaluation.
Dr. Lee's current research interests are network science, game theory,
machine learning and high-speed networks.  He is a senior IEEE member.
\end{IEEEbiography}

\begin{IEEEbiography}[{\includegraphics[width=1in,height=1.25in,clip,keepaspectratio]{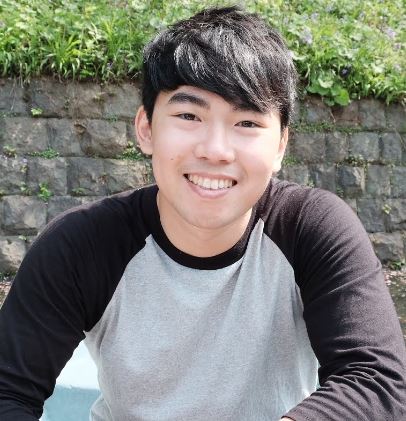}}]
{Yu Lun Lin} received his B.S. degree in electrical engineering from National Tsing-Hua University, Hsinchu, Taiwan, in 2016. He is currently pursuing the M.S. degree in the Institute of Communications Engineering, National Tsing-Hua University. His research interest is in multichannel rendezvous problems in cognitive radio networks.
\end{IEEEbiography}

\begin{IEEEbiography}[{\includegraphics[width=1in,height=1.25in,clip,keepaspectratio]{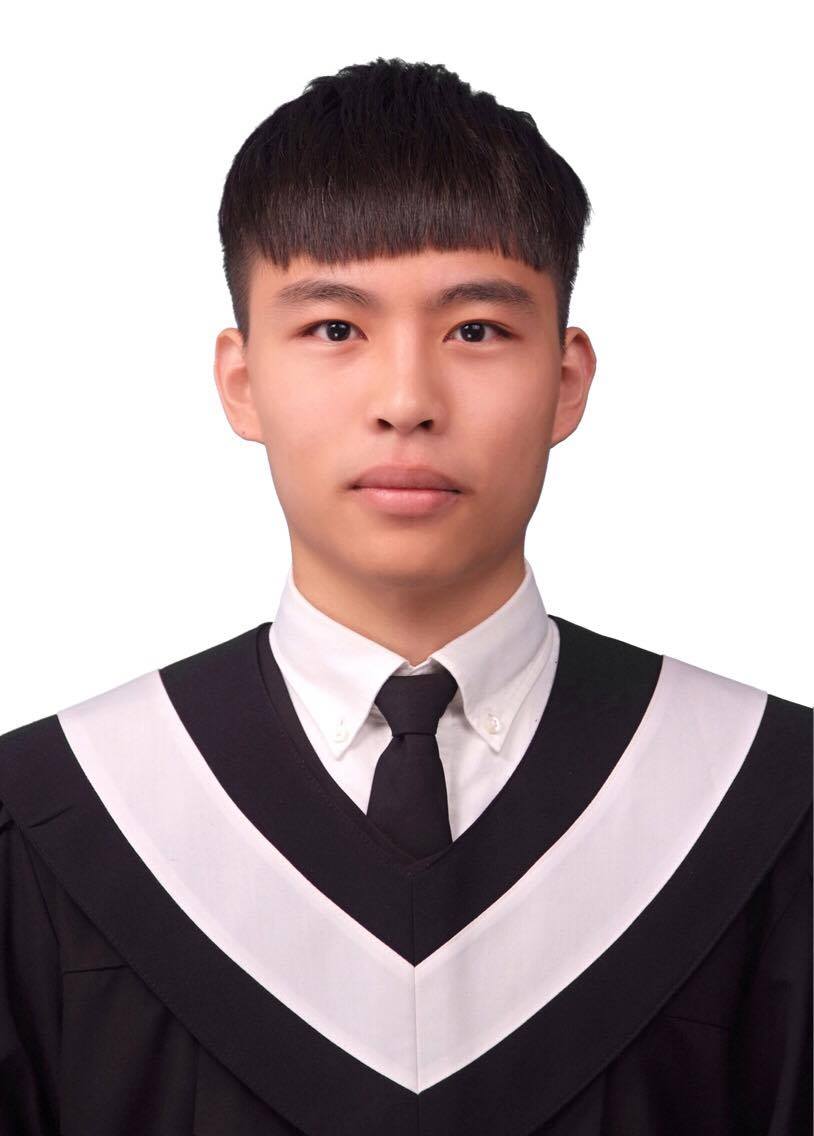}}]
 {Jen-Hung Wang} received his B.S. degree in electrical
engineering from Chang Gung University, Taoyuan,
 Taiwan, in 2017. He is currently pursuing
the M.S. degree in the Institute of Communications
Engineering, National Tsing-Hua University.
His research interest is in multichannel rendezvous
problems in cognitive radio networks.
\end{IEEEbiography}

\end{document}